\documentclass[twocolumn,showpacs,noshowkeys,floatfix,prd,aps]{revtex4}
\usepackage{epsfig} 
\usepackage{longtable} 
\usepackage{hyperref}
\usepackage{bm}
\begin{document}  
\begin{titlepage} 
\null 
\vspace{2cm} 
\begin{center} 
\Large\bf  
Search for the lepton flavor violating decay \bm{$A^0/H^0\to\tau^\pm\mu^\mp$} 
at hadron colliders
\end{center} 
\vspace{1.5cm} 
 
\begin{center} 
\begin{large} 
K\'et\'evi~Adikl\`e~Assamagan\\ 
\end{large} 
\vspace{0.5cm} 
Department of Physics, Brookhaven National Laboratory\\ 
Upton, NY 11973 USA\\ 
\vspace{0.7cm}  
\begin{large} 
Aldo Deandrea and Pierre-Antoine Delsart\\ 
\end{large} 
\vspace{0.5cm} 
Institut de Physique Nucl\'eaire, Universit\'e de Lyon I\\  
4 rue E.~Fermi, F-69622 Villeurbanne Cedex, France 
\end{center} 
 
\vspace{1.3cm} 
 
\begin{center} 
\begin{large} 
{\bf Abstract}\\[0.5cm] 
\end{large} 
\parbox{14cm}
{ 
In the two Higgs doublet model type III and in several other 
extensions of the Standard Model, there are no discrete symmetries 
that suppress flavor changing couplings at tree level. The experimental 
observation of the $\nu_\mu$--$\nu_\tau$ flavor oscillation may suggest the 
non-conservation of lepton number. This would lead to the decay of the type 
$A^0/H^0 \to \tau^\pm\mu^\mp$. We determine the present low energy limit on 
lepton 
flavor violating (LFV) couplings from the muon $g-2$ measurement and 
discuss the prospects for detecting lepton flavor violating decays at the 
TeVatron and at the Large Hadron Collider. The achievable bounds on the LFV 
coupling parameter $\lambda_{\tau\mu}$ are presented.
}  
\end{center}   
\vspace{2cm} 
\noindent 
PACS: 12.60.Fr, 11.30.Hv, 14.80.Cp\\ 
\vfil 
\noindent 
LYCEN-2002-27\\ 
BNL-69300\\
ATL-COM-PHYS-2002-031\\
July 2002\\ 
\vfill 
\eject 
\end{titlepage} 
 
\newpage 
%%%%%%%%%%%%%%%%%%%%%%%%%%%%%%%%%%%%%%%%%%%%%% 
 
\title{Search for the lepton flavor violating decay 
\bm{$A^0/H^0\to\tau^\pm\mu^\mp$} at 
hadron colliders}  
\author{K\'et\'evi~A. Assamagan}  
\email{ketevi@bnl.gov} 
\affiliation{Department of Physics, Brookhaven National Laboratory, 
Upton, NY 11973 USA}
\author{Aldo Deandrea}
\email{deandrea@ipnl.in2p3.fr}
\affiliation{Institut de Physique Nucl\'eaire, Universit\'e Lyon I, 4 rue
E.~Fermi,  F-69622 Villeurbanne Cedex, France}   
\author{Pierre-Antoine Delsart}
\email{delsart@ipnl.in2p3.fr}
\affiliation{Institut de Physique Nucl\'eaire, Universit\'e Lyon I, 4 rue
E.~Fermi,  F-69622 Villeurbanne Cedex, France}   
\date{July, 2002}  
\preprint{LYCEN-2002-27}
\preprint{BNL-69300} 
\preprint{ATL-COM-PHYS-2002-031}
\pacs{12.60.Fr, 11.30.Hv, 14.80.Cp} 
\keywords{2HDM-III; Lepton Flavor Violation; Higgs boson}

\def\OOrd{\lower .7ex\hbox{$\;\stackrel{\textstyle >}{\sim}\;$}}
\def\OOle{\lower .7ex\hbox{$\;\stackrel{\textstyle <}{\sim}\;$}}

\begin{abstract}  
In the two Higgs doublet model type III and in several other 
extensions of the Standard Model, there are no discrete symmetries 
that suppress flavor changing couplings at tree level. The experimental 
observation of the $\nu_\mu$--$\nu_\tau$ flavor oscillation may suggest the 
non-conservation of lepton number. This would lead to the decay of the type 
$A^0/H^0 \to \tau^\pm\mu^\mp$. We determine the present low energy limit on 
lepton 
flavor violating (LFV) couplings from the muon $g-2$ measurement and 
discuss the prospects for detecting lepton flavor violating decays at the 
TeVatron and at the Large Hadron Collider. The achievable bounds on the LFV 
coupling parameter $\lambda_{\tau\mu}$ are presented.
\end{abstract}  
 
\maketitle  

\section{Motivation}
\label{sec:motivation}
In the Standard Model (SM), lepton flavor is conserved separately for each 
generation. The diagonalization of the up-type and down-type mass matrices 
ensures the diagonalization of the Higgs-fermion coupling 
matrices~\cite{2HDM}: the interaction term of the neutral fields in the SM can 
be written as:
\begin{equation}
\label{eq:Yint}
\mathcal{L}_Y = -h_{ij} \bar{\psi}_i\psi_j\phi.
\end{equation}
The spontaneous electro-weak symmetry breaking gives the mass matrix
\begin{equation}
\label{eq:mat}
M_{ij} = h_{ij}<\phi>.
\end{equation}
Diagonalizing $M_{ij}$ also diagonalizes the Yukawa coupling matrix $h_{ij}$. 
The severe experimental limits on the existence of flavor changing neutral 
currents place stringent constraints on the flavor changing sector of 
extended models~\cite{FCNC} where lepton flavor violation (LFV) may appear at
tree  level or may be induced at higher orders. In the Minimal Supersymmetric
Standard  Model  (MSSM) the flavor problem is related to the soft
supersymmetry breaking mass  terms. 
In the basis where the lepton mass matrix is diagonalized, if there are 
non-zero  off-diagonal matrix elements in the slepton mass matrix, LFV is 
introduced 
via loop contributions involving slepton mixing. There are many ways to avoid 
LFV, for example gravity \cite{Nilles:1984ge} or gauge mediated~\cite{giuratt} 
supersymmetry breaking, or flavor symmetries~\cite{Barbieri:1996uv}.
In the minimal super-gravity model (SUGRA) the supersymmetry breaking mass 
terms have a universal structure at a high scale of the order of the Planck 
scale. However LFV effects can be induced by radiative 
corrections~\cite{Hall:1986dx}. Large LFV effects can arise in 
supersymmetric models  (SUSY) with a  right-handed Majorana 
neutrino~\cite{Borzumati:1986qx,JELL,FENG,HISA} and in SUSY with R-parity 
violation~\cite{Huitu}.

In general, in models with several Higgs doublets, the up-type quarks and the 
down-type quarks can simultaneously couple to more than a single scalar 
doublet. As a result, the same operators do not diagonalize the mass matrices 
and the Higgs-fermion couplings, leading to the prediction of Flavor Changing 
Neutral Current (FCNC) at tree level. For instance in the two-Higgs Doublet 
Model (2HDM), the Yukawa interaction Lagrangian (for the neutral fields) 
can be written as:
\begin{equation}
\label{eq:lag}
 \mathcal{L}_Y = -f_{ij}\bar{\psi}_i\psi_j\varphi_1-
g_{ij}\bar{\psi}_i\psi_j\varphi_2
\end{equation}
which gives, after spontaneous electro-weak symmetry breaking, a mass matrix 
of the form:
\begin{equation}
\label{eq:2mat}
M_{ij} = f_{ij} <\varphi_1> + g_{ij} <\varphi_2>.
\end{equation}
When this matrix $M_{ij}$ is diagonalized, the coupling matrices $f_{ij}$ 
and $g_{ij}$ are not, in general, diagonalized. To suppress tree level 
FCNC in the theory so as not to be in conflict with known 
experimental limits, an \textit{ad hoc} discrete symmetry is 
invoked~\cite{GLAS-WEIN} whereby the fermions of a given electric charge 
could couple to no more than one Higgs doublet. In the 2HDM, the up-type and 
the down-type quarks couple either to the same Higgs doublet (this is known 
as the 2HDM-I), or they could couple to different doublets (2HDM-II). One of 
the most stringent test of the 2HDM type I and type II comes from the 
measurement of 
the $b \to s\gamma$ decay rate which receives substantial enhancement (over 
the SM prediction) in the 2HDM in a large region of 
the ($m_{H^\pm}$, $\tan\beta$) parameter space~\cite{BWOS,BCH,CHET}. The 
measured $b \to s\gamma$ decay rate from CLEO~\cite{CLEO} and 
ALEPH~\cite{ALEPH} leads to a model dependent indirect lower bound of the 
charged Higgs mass as function of $\tan\beta$~\cite{HP-BOUND}.        

In the 2HDM-III, no discrete symmetries are present and in general FCNC exist 
in this model~\cite{ATWO,DIAZ}. As an example the LFV interaction Lagrangian 
of the light neutral Higgs boson $h$ of the 2HDM-III type b (see the 
Appendix for details) is:
\begin{equation} 
\label{eq:fc_lag}
-\mathcal{L}_{LFV} =h_{ij} \; \bar{l}_il_jh + {\mathrm {h.c.}} = 
\xi_{ij}\frac{\cos(\alpha-\beta)}{\sqrt{2}\cos\beta} \; \bar{l}_il_jh + 
{\mathrm {h.c.}} 
\end{equation}
where $\alpha$ is the mixing angle of the neutral Higgs sector and 
$\xi_{ij}$ the Yukawa LFV couplings and $i$, $j$ 
are the generation indices (in the following the notation $h_{ij}$ will be 
used to indicate the generic Yukawa coupling including the mixing angles). To 
be consistent with experimental data on 
$K^0-\bar{K}^0$, $D^0-\bar{D}^0$ and $B^0-\bar{B}^0$ mixing which put 
stringent constraints on flavor changing couplings with the first generation 
index, and since one might expect the biggest contribution to come from the 
LFV couplings of the second and the third generation 
($\xi_{sb}$, $\xi_{\tau\mu}$, $\xi_{ct}$), these couplings have 
been parameterized as a function of the masses of the fermions involved since 
a natural hierarchy is found in the fermion masses~\cite{FCNC}:
\begin{equation}
\label{eq:coupl}
\xi_{ij} = \lambda_{ij}\frac{\sqrt{m_im_j}}{v},
\end{equation}
where $v \simeq 246$~GeV and the residual arbitrariness of flavor changing 
couplings is expressed by the parameters $\lambda_{ij}$ which is 
constrained by experimental bounds on FCNC and LFV processes. A similar 
hierarchy will be assumed for the LFV coupling $\eta_{ij}$ of 2HDM-III type a 
--- 
see the appendix. In the charged Higgs decays this implies a zero LFV coupling 
if the neutrino is massless and in general a suppression proportional to the 
square root of the small neutrino mass. We also consider an alternative case in 
which we drop the neutrino mass dependence for the charged Higgs LFV couplings 
and adopt instead the same parameterization as in the neutral Higgs sector. In 
the numerical analysis of the muon anomalous magnetic moment, this distinction 
is not important as the charged Higgs contribution is small in both cases and 
can be neglected in comparison to the neutral ones for the range of masses and 
mixing angles considered. 

\section{Low energy bounds}
\label{sec:lowen}

\begin{figure}
\epsfxsize=7truecm
\begin{center}
\epsffile{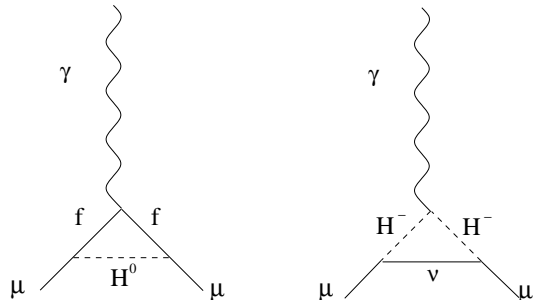}
\caption{The one-loop contributions of the Higgs sector to $a_\mu$. $H^0$ 
stands for a generic neutral Higgs boson, $f$ is a lepton. With $f=\mu$ we 
obtain the flavor conserving contribution, with $f=\tau$ the LFV one. As 
explained in the text we neglect the LFV contribution with $f=e$.}
\label{fig:oneloop}
\end{center}
\end{figure}

\begin{figure}
\epsfxsize=8truecm
\begin{center}
\epsffile{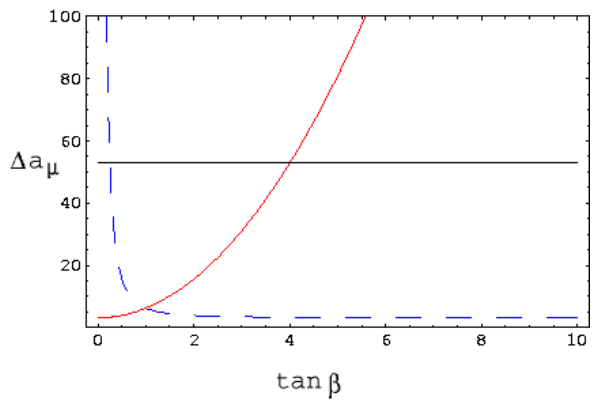}
\caption{$\Delta a_\mu$ in units of $10^{-10}$ as a function of $\tan\beta$ 
using set 2 of parameters for the 2HDM-III type a (dashed line) and 2HDM-III
type b (continuous line).  The region above the horizontal line is excluded
at 90 \% CL by the muon $g-2$  data.}
\label{fig:limtgb}
\end{center}
\end{figure}

In the purely leptonic sector, the $\mu \to e\gamma$ conversion process gives 
$\sqrt{\lambda_{\tau\mu}\lambda_{e\tau}} < 5$~\cite{SHER3}. It would be 
desirable to examine a process that depends only on a single coupling. One 
such process is the muon anomalous magnetic moment $a_\mu=(g_\mu-2)/2$ 
\cite{BOUND} where high precision data~\cite{Brown:2001mg} can be used to 
constrain $\lambda_{\tau\mu}$, by comparing the measured $a_\mu$ to the 
theoretical prediction of the SM. The new experimental world average reads 
~\cite{Brown:2001mg}:
\begin{equation}
a_\mu^{\mathrm{exp}}= 11659203(8) \times 10^{-10}.
\end{equation}
Recently, the standard model calculation was revised in order to take into 
account the correct sign for the light by light hadronic contribution 
\cite{llh} and the standard model expectation is :
\begin{equation}
a_\mu^{\mathrm{SM}}= 11659177(7) \times 10^{-10}.
\label{eq:th1}
\end{equation}
Another often quoted value is 
\begin{equation}
a_\mu^{\mathrm{SM}}= 11659186(8) \times 10^{-10}
\label{eq:th2}
\end{equation}
which gives only slightly more restrictive figures if used to bound the LFV 
couplings. In the following we shall use the value of Equation~(\ref{eq:th1}). 
Note that a recent evaluation of the light by light hadronic correction 
\cite{Ramsey-Musolf:2002cy} based on chiral perturbation theory suggests that 
the theoretical error due to unknown low energy constants from sub-leading 
contributions may increase the estimated error. The difference between 
experiment and the SM theoretical calculation is:
\begin{equation}
\Delta a_\mu=a_\mu^{\mathrm{exp}}-a_\mu^{\mathrm{SM}}= 26(11) \times 
10^{-10}.
\end{equation} 
We obtain the 90\% confidence level (CL) range on $\Delta a_\mu$ 
\begin{equation}
\label{eq:da_bounds}
8 \times 10^{-10} \le\Delta a_\mu \le 44 \times 10^{-10}
\end{equation} 
to constrain new physics. In the following we shall consider the effect of 
flavor violating Higgs-leptons interactions plus the flavor conserving 
Higgs bosons contributions as the only additional ones with respect to the SM.
At one-loop level the Feynman diagrams are those of  Figure~\ref{fig:oneloop} 
and the contribution to $a_\mu$ is given for a large class of models by 
\cite{Leveille:1978rc} (see also the erratum in \cite{2HDM} concerning other 
results in the literature), and can be
used to obtain the one-loop Higgs contributions to $a_\mu$ for the model 
considered 
in this paper:
\begin{equation}
\Delta a_\mu^N = \frac{h^2_{\mu f} m_\mu^2}{8 \pi^2} \int^1_0 \frac{x^2 
(1-x) \pm x^2 (m_f/m_\mu)}{m_\mu^2 x^2 +x (m_f^2-m_\mu^2) +(1-x) m^2_H }dx
\end{equation}
for a neutral Higgs boson and the sign is $+(-)$ for the scalar 
(pseudo-scalar). $h_{ij}$ is here a generic Yukawa coupling, whose expression 
in terms of $\eta_{ij}$ or $\xi_{ij}$ and the angles $\alpha$, $\beta$ can be 
read in the Lagrangian given in the appendix. $m_f$ is the mass of the muon
for the  flavor conserving contribution with coupling $h_{\mu \mu}$, and 
$m_f=m_\tau$ 
for the LFV contribution with coupling $h_{\mu \tau}$. We neglect the electron 
contribution as the coupling $h_{\mu e}$ is more constrained and because of 
the natural 
hierarchy assumed in formula (\ref{eq:coupl}). For the charged Higgs boson 
we have the same coupling for the scalar and pseudoscalar 
contributions in the Lagrangian, therefore we give the sum of the two 
in one formula:
\begin{equation}
\Delta a_\mu^C = \frac{h^2_{\mu\nu} m_\mu^2}{8 \pi^2} \int^1_0 \frac{2 
x^2 (x-1)}{m_\mu^2 x^2 +x (m^2_H-m^2_\mu)}dx
\end{equation}
where we neglected terms proportional to the neutrino mass.
In order to give the bounds coming from the $g-2$ measurement we choose
the sets of mass and mixing angle parameters of Table~\ref{tab:set}.
\begin{table}[h]
\begin{center}
\begin{minipage}{0.9\linewidth} 
\caption{\label{tab:set}The four sets of parameters used to obtain bounds on 
the LFV couplings. Set 1--3 are consistent with the relations between 
masses and mixing angles obtained at one-loop within the MSSM 
\cite{Haber} in order to allow for a comparison. Note however that the  
2HDM-III is not constrained by the symmetries imposed on MSSM in order to 
avoid tree-level LFV. Set 4 corresponds to a choice of parameters that is not 
allowed in MSSM. Masses are in GeV and the angle $\alpha$ in rad.}
\end{minipage}
\vbox{\offinterlineskip 
\halign{&#& \strut\quad#\hfil\quad\cr  
\colrule
& Set & $m_h$ &  $m_H$ & $m_A$ & $m_{H^\pm}$ & $\alpha$ & $\tan \beta$ & 
\cr \colrule
& \textbf{(1)} & 93 & 134 & 100 & 127 & 0.4 & 5 & \cr
& \textbf{(2)} & 127 & 131  & 129 & 160 & -0.58 & 45 & \cr 
& \textbf{(3)} & 128 & 500  & 496 & 509 & 0 & 50 & \cr 
& \textbf{(4)} & 125 & 200  & 200  & 250 & 0.2 & 10 & \cr 
\colrule}}
\end{center}
\end{table}
By calculating the contribution to $a_\mu$ from the Higgs sector we obtain 
limits on the LFV couplings of 2HDM-III type a and b --- we use only the upper 
limits of~(\ref{eq:da_bounds}) to derive the muon $g-2$ bounds on the LFV couplings. 
The results are in Table~\ref{tab:limit}.
\begin{table}[h]
\begin{center}
\begin{minipage}{0.9\linewidth} 
\caption{\label{tab:limit}The 90\% CL limits on the LFV couplings  
$\lambda_{\tau\mu}$, $\xi_{\tau\mu}$, $\eta_{\tau\mu}$ from the experimental 
measurement of $a_\mu$.}
\end{minipage}
\vbox{\offinterlineskip 
\halign{&#& \strut\quad#\hfil\quad\cr  
\colrule
& Set & 2HDM-III type a &  2HDM-III type b &
\cr \colrule
& \textbf{(1)} & $\lambda_{\tau\mu} < 31$ ($\eta_{\tau\mu} < 0.06$)& 
$\lambda_{\tau\mu} < 6.3$ ($\xi_{\tau\mu} < 0.012$)&  \cr
& \textbf{(2)} & $\lambda_{\tau\mu} < 38$ ($\eta_{\tau\mu} < 0.07$)& 
$\lambda_{\tau\mu} < 0.8$ ($\xi_{\tau\mu} < 0.002$)&  \cr
& \textbf{(3)} & $\lambda_{\tau\mu} < 123$ ($\eta_{\tau\mu} < 0.24$)& 
$\lambda_{\tau\mu} < 2.5$ ($\xi_{\tau\mu} < 0.005$)&  \cr 
& \textbf{(4)} & $\lambda_{\tau\mu} < 53$ ($\eta_{\tau\mu} < 0.10$)& 
$\lambda_{\tau\mu} < 5.3$ ($\xi_{\tau\mu} < 0.010$)&  \cr 
\colrule}}
\end{center}
\end{table}
In Figure~\ref{fig:limtgb} we show the values of $\Delta a_\mu$ given by the 
2HDM-III using the set $\textbf{(2)}$ of parameters of Table~\ref{tab:set} 
with $\lambda_{\mu\nu}=10$ as 
a function of $\tan \beta$. In model type a, $\Delta a_\mu$ is almost flat for 
$\tan \beta >2$, while in model type b it is a growing function of 
$\tan\beta$. The same is true for the other sets of parameters. In both models
the Higgs sector contribution to $\Delta a_\mu$ is a growing function of the 
LFV couplings.

\section{Collider experiments}
\label{sec:colliders}

Bounds on other LFV couplings $\lambda_{ij}$ can be obtained from the 
processes 
$B_s \to \mu\mu$, $B \to X_s\mu\mu$, $e^+e^- \to \bar{t}c\nu_e\bar{\nu}_e$,
$e^+e^- \to tt\bar{c}\bar{c}$, $\mu^+\mu^- \to tc$, $B_s \to (K)\tau\mu$ and 
$B_s \to (K)\tau\tau$~\cite{ATWO}. However, the bounds obtained from these 
processes 
would involve two couplings as in the case of $\mu \to e\gamma$. The flavor 
changing process
$t \to H^0 c$ has been extensively studied for the LHC~\cite{AGUI}; in the 
context of the 2HDM-I and II, top quark decays beyond the SM,
\begin{equation}
\label{eq:top-decay}
t \to ch\,\, (h = h^0, H^0, A^0),
\end{equation}
were studied in~\cite{SANT} and it is shown that these processes could be 
accessible at the 
LHC and at the linear collider; the prospects for detecting the decay $t \to 
cH$ at the 
$e^+e^-$ linear collider have also been investigated in~\cite{Han-Sher}.
 
The $\nu_\mu - \nu_\tau$ flavor mixing observed in the atmospheric neutrino 
experiments~\cite{KAMIO} would lead to the flavor violating decays
\begin{eqnarray}
\label{eq:t-->mugamma}
\tau^\pm & \to & \mu^\pm\gamma, \\
\label{eq:t-->mumumu}
\tau^\pm & \to & \mu^\pm\mu^+\mu^-, \\
\label{eq:chi2 -->chi1taumu}
\tilde{\chi}^0_2 & \to & \tilde{\chi}^0_1\tau\mu, \\
\label{eq:h-->taumu}
h & \to & \tau^\pm \mu^\mp.
\end{eqnarray}
SUSY can accommodate the observed flavor mixing~\cite{JELL,FENG,HISA} and 
thus, the LFV processes~(\ref{eq:t-->mugamma}), (\ref{eq:t-->mumumu}) and 
(\ref{eq:chi2 -->chi1taumu}) would arise in these models. A study conducted at 
the LHC showed that an upper bound of $0.6\times 10^{-6}$ on the $\tau^\pm \to 
\mu^\pm\gamma$ branching ratio can be 
achieved with an integrated luminosity of 30 fb$^{-1}$ while theoretical 
estimates are at the level of $10^{-9}$ or less~\cite{HISA,SERI}. Direct 
evidence 
of LFV in the slepton sector of SUSY would be inferred in the observation of 
the process~(\ref{eq:chi2 -->chi1taumu}) which has 
also been studied for the LHC~\cite{HINC}. It was shown that in some cases, 
the direct evidence would offer better sensitive than the $\tau^\pm  \to  
\mu^\pm\gamma$ process. 

The decay $h \to \tau^\pm\mu^\mp$ can be accommodated in the 2HDM-III where 
no discrete symmetry suppresses the LFV couplings at tree level, and the 
partial decay width is parameterized by the LFV coupling $\lambda_{\tau\mu}$. 
The decay $H^0 \to l_i^+ l_i^-$ (SM-like) will be used in the following 
as a comparison for the LFV decays. Its partial width is 
\begin{equation}
\Gamma_{SM}(H^0 \to l_i^+ l_i^-) = m_H 
\frac{1}{8\pi}\frac{m_i^2}{v^2}\; ,
\label{eq:smdec}
\end{equation}
where we neglect small terms of the type $m_i/m_H$ --- see the appendix for 
complete expressions. The partial width of the decay $H^0 \to l_i^\pm l_j^\mp$ 
(where $l=e$, $\mu$, $\tau$ and $i\neq j$) is for 2HDM-III type a
\begin{equation}
\label{eq:h-to-ll-1}
\Gamma(H^0 \to l_i^\pm l_j^\mp) = m_H 
\frac{\lambda^2_{ij}}{8\pi}\frac{m_im_j}{v^2}\frac{\sin^2(\alpha-\beta)}{2 
\sin^2 \beta}
\end{equation}
and 
\begin{equation}
\label{eq:h-to-ll-2}
\Gamma(H^0 \to l_i^\pm l_j^\mp) = m_H 
\frac{\lambda^2_{ij}}{8\pi}\frac{m_im_j}{v^2}\frac{\sin^2(\alpha-\beta)}{2 
\cos^2 \beta}
\end{equation}
for 2HDM-III type b.

Hadron colliders may be sensitive to the processes $h \to e^\pm\mu^\mp$ and 
$h \to \tau^\pm e^\mp$~\cite{TOSC}, particularly at high luminosity, but 
these decays are not considered in the present study which is 
further motivated by a favorable interpretation of the atmospheric neutrino 
mixing experiments. It is shown in~\cite{SHER2} that the muon collider would be
sensitive to $H^0 \to \tau^\pm\mu^\mp$. The non-observation of this process 
for $m_H < 140$~GeV at the muon collider in addition to the failure to detect 
the top quark decay $t\to cH^0$ at the LHC~\cite{AGUI} would rule out the 
2HDM-III~\cite{SHER2}.

In this paper, we present the prospects for the detection of the LFV decay 
$A^0/H^0 \to \tau^\pm \mu^\mp$ at the LHC and TeVatron. We shall consider the 
2HDM-III and we shall parameterize  the $A^0/H^0 \to \tau^\pm \mu^\mp$ 
branching ratio ($BR$) by the LFV coupling parameter $\kappa_{\tau\mu}$ 
\cite{Cotti:2001fm} with respect to the SM-like decay $H^0 \to \tau^+ \tau^-$ 
given in formula (\ref{eq:smdec}):
\begin{equation}
\label{eq:br-taumu}
BR (A^0/H^0 \to \tau\mu) = \kappa^2_{\tau\mu} 
\left(\frac{2m_\mu}{m_\tau}\right)
BR_{SM}(H^0 \to \tau\tau)
\end{equation}
where the dependence on $\alpha$, $\beta$, the ratio of the total widths and
$\lambda_{\tau\mu}$ is absorbed  into the LFV coupling parameter
$\kappa_{\tau\mu}$. For example, for the decay  and the model considered in
formula (\ref{eq:h-to-ll-1}) we have: \begin{equation}
\label{eq:kl1}
\kappa_{\tau\mu} = \lambda_{\tau\mu}\frac{\sin(\alpha-\beta)}{\sqrt{2} 
\sin\beta} \; \sqrt{\frac{\Gamma_T^{SM}}{\Gamma_T^{a}}}
\end{equation}
while from formula (\ref{eq:h-to-ll-2}) we obtain:
\begin{equation}
\label{eq:kl2}
\kappa_{\tau\mu} = \lambda_{\tau\mu}\frac{\sin(\alpha-\beta)}{\sqrt{2} 
\cos\beta}\; \sqrt{\frac{\Gamma_T^{SM}}{\Gamma_T^{b}}}
\end{equation} 
where $\Gamma_T^{SM}$ is the total SM-like width and $\Gamma_T^{a,b}$
is the total width in model $a$, $b$ respectively. Similar formulas can be 
written for the $h^0$ and $A^0$ Higgs bosons. For $h^0$ one has to replace 
$\sin (\alpha-\beta)$ in equations (\ref{eq:kl1}) and (\ref{eq:kl2}) with 
$\cos (\alpha-\beta)$. For $A^0$ one has to replace 
$\sin (\alpha-\beta)$ with $1$.

In Table~\ref{tab:k-l} we give examples of the correspondence between the 
parameterization in terms of $\lambda$ and the one in terms of $\kappa$ for 
set~1 and set~2 of Table~\ref{tab:set}.

\begin{table}[h]
\begin{center}
\begin{minipage}{0.9\linewidth} 
\caption{\label{tab:k-l}The correspondence between the parameters $\kappa$ and 
$\lambda$ using set~1 (set~2) of Table~\ref{tab:set} for the LFV couplings of 
the Higgs bosons $H^0$ 
and $A^0$.}
\end{minipage}
\vbox{\offinterlineskip 
\halign{&#& \strut\quad#\hfil\quad\cr  
\colrule
& & $\lambda=1$ &  $\lambda=5$ & $\lambda=10$ & 
\cr \colrule
& type a &  &  &  & \cr \colrule
& $\kappa (H^0)$ & 1.1 (1.2) & 9.4 (6.2) & 7.8 (12.2)&\cr 
& $\kappa (A^0)$ & 3.3 (30.7) & 16.6 (68) & 7 (72.6) &\cr \colrule
& type b &  &  &  & \cr \colrule
& $\kappa (H^0)$ & 1.1 (0.7) & 5 (3.3) & 7.6 (6.5)&\cr 
& $\kappa (A^0)$ & 0.1 (0.001) & 0.5 (0.006) & 1 (0.01) &\cr 
\colrule}}
\end{center}
\end{table}
We shall discuss the achievable bounds on $\kappa_{\tau\mu}$ and 
$\lambda_{\tau\mu}$ in the following sections. 

\section{Search for \bm{$A^0/H^0 \to\tau\mu$}}
\label{sec:lhc}

\begin{figure}
\epsfxsize=3.5truecm
\begin{center}
\epsffile{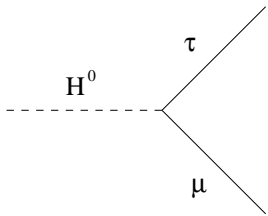}
\caption{The Higgs decay through the lepton flavor violating coupling 
$H^0\tau\mu$.}
\label{fig:htaumu}
\end{center}
\end{figure}

\begin{figure}
\epsfxsize=8truecm
\begin{center}
\epsffile{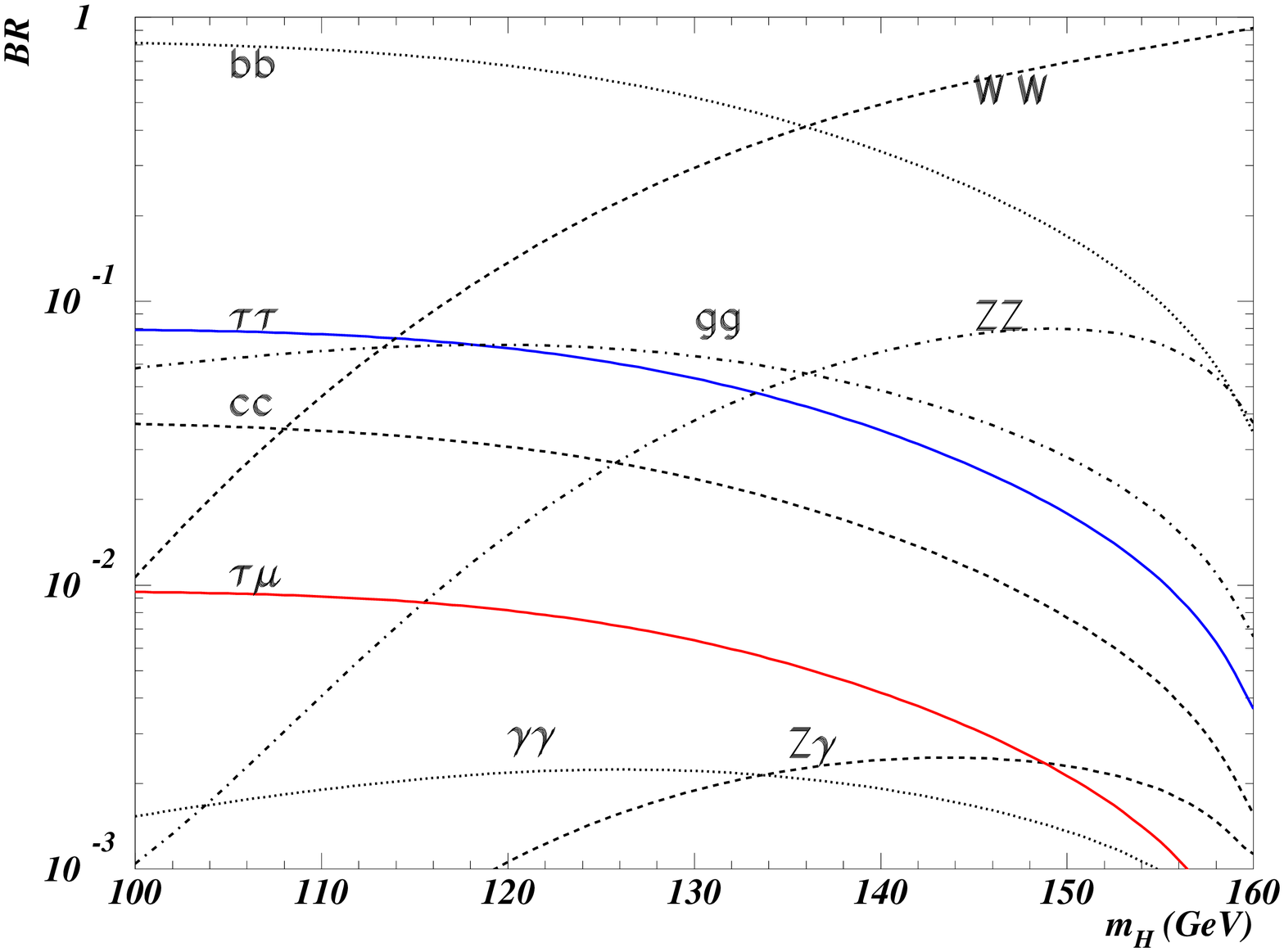}
\caption{The Higgs boson decay branching ratios as a function of $m_H$. For 
the 
$A^0/H^0 \to \tau^+\mu^- + \tau^-\mu^+$ channel, the coupling parameter 
$\kappa_{\tau\mu}$ is taken to be one.}
\label{fig:htaumu_1}
\end{center}
\end{figure}

We consider the production of the neutral Higgs bosons $A^0$ and $H^0$ 
through gluon fusion, 
\begin{figure}
\epsfxsize=6truecm
\begin{center}
\epsffile{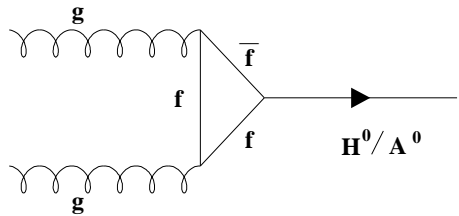}
\caption{Higgs boson production mechanism through gluon fusion.}
\label{fig:ggh}
\end{center}
\end{figure}
$gg \to A^0/H^0$ (see Figure~\ref{fig:ggh}), and the LFV decay 
$A^0/H^0 \to \tau^\pm\mu^\mp$ (Figure~\ref{fig:htaumu}). We restrict the 
present work to the low mass region, $120 < m_A < 160$~GeV, primarily because 
the SM decay $H^0_{SM} \to \tau^+\tau^-$, hence $ A^0/H^0\to \tau^+\mu^-$ --- 
see Equation~(\ref{eq:br-taumu}) --- becomes negligible~\cite{HDECAY} as the 
SM mode $H^0_{SM} \to W^+W^-$ opens up around 160~GeV as shown in 
Figure~\ref{fig:htaumu_1} where we assume $\kappa_{\tau\mu}=1$. We take as a
reference the parameters of set~2 in Table~\ref{tab:set} for comparison 
with the MSSM case without loss of generality.
The events are generated in PYTHIA6.2~\cite{PYTHIA} with CTEQ5L~\cite{CTEQ} 
parton 
distribution function parameterization, and with the detector resolution and 
efficiencies  parameterization of ATLFAST~\cite{ATLFAST} from full detector 
simulations. 

We search for a final state where the $\tau$ lepton decays to hadrons, 
$\tau\to\mbox{jet}\,\nu_\tau$ with a branching ratio of $\sim 65$\% or 
to an electron, $\tau\to e\nu_e\nu_\tau$ ($BR \sim 18$\%). The main 
backgrounds --- in both final states --- include the $W^+W^-$ pair production, 
the Drell-Yan type process $Z^0(\gamma^*) \to \tau^+\tau^-$. For the hadronic 
$\tau$ decays, an additional background comes from $W^\pm +$jets events where 
a jet is mis-identified as a $\tau$ jet:
\begin{eqnarray}
\label{eq:backg}
pp(\bar{p}) \to & W^\pm Z^0 & \to \mu^\pm\nu_\mu\tau^+\tau^-, \\
   \to & W^+ W^- & \to \mu^+\nu_\mu\tau^-\bar{\nu}_\tau, \nonumber \\
   \to & t\bar{t}& \to \mu^\pm\nu_\mu b\tau^\mp\nu_\tau\bar{b}, \nonumber \\
   \to & Z^0(\gamma^*) \to \tau^+\tau^- & \to 
\mu^+\nu_\mu\bar{\nu}_\tau\tau^-, \nonumber \\
   \to & W^\pm+\mbox{jets} & \to \mu^\pm\nu_\mu+\mbox{jets}. \nonumber
\end{eqnarray}
The $gg \to A^0/H^0$ cross sections are calculated using the program 
HIGLU~\cite{HIGL}. 
The signal cross-sections have been calculated at next-to-leading order (NLO) 
and next-next-to-leading order (NNLO)~\cite{HIGL,BALA}. For the backgrounds, 
NLO estimates are available~\cite{CATA,FRIX,DIXO}, except for $W^\pm +$jets 
where 
NLO calculations have been performed for a vector boson production with 2~jets 
at the TeVatron~\cite{CAMP}. We have therefore used the leading order (LO) 
estimates of the signal and background cross sections. 

Unless explicitly stated otherwise, the normalizations of the figures 
referenced in the sections~\ref{sec:hadronic}--\ref{sec:lep_decay} are that 
of three years at low luminosity for one experiment at the LHC using the 
rates shown in Table~\ref{tab:htaumu_1}.
\begin{table*}
\begin{center}
\begin{minipage}{.75\linewidth} 
\caption{\label{tab:htaumu_1} The rates, $\sigma\,\times\,$~BR(pb), for the 
signal $gg\to A^0/H^0 \to \tau^+\mu^- + \tau^-\mu^+$, and the  backgrounds at 
the LHC. The dominant backgrounds are $Z \to \tau\tau$ and $W^\pm 
+\mbox{jets}$ 
where $W^\pm \to \mu^\pm\nu_\mu$ and a jet is mis-identified as a $\tau$ jet. 
We assume the coupling parameter $\kappa_{\tau\mu} = 1$ and $\tan\beta=45$ in 
the estimate of the signal rates. An additional background comes 
$gg\to A^0/H^0 \to \tau^+\tau^-$ with one $\tau$ decaying to $\mu$, 
$\tau \to \mu\nu_\mu\nu_\tau$ and the other $\tau$ decays to hadrons. At 
$\tan\beta=45$, the scalar and the pseudo-scalar Higgs bosons are degenerate 
in mass for $m_A \geq 130$~GeV, and the relative strengths of 
$gg\to A^0\to\tau\mu$ and $gg\to H^0\to\tau\mu$ are not important. 
For $m_A=120$~GeV, the $A^0$ and the $H^0$ bosons have comparable strengths.}
\end{minipage}
\vbox{\offinterlineskip 
\halign{&#& \strut\quad#\hfil\quad\cr  
\colrule
& Process   && $m_A$ (GeV)  && $m_H$ (GeV) && $\sigma\,\times\,$~BR(pb)   & \cr
\colrule
& $gg\to A^0/H^0 \to \tau^+\mu^- + \tau^-\mu^+$                            && 
$119.3$  && 128.4    &&   7.5     & \cr
&                                                                      && 
$129.3$  && 130.1    &&   4.5      & \cr
&                                                                      && 
$139.2$  && 140.2    &&   2.1      & \cr
&                                                                      && 
$149.1$  && 150.0     &&   0.8      & \cr
&                                                                      && 
$159.1$  && 160.0     &&   0.1       & \cr
\colrule
\colrule
& $gg\to A^0/H^0 \to \tau\tau$                                         && 
$119.3$    && 128.4  &&   99.5     & \cr
&                                                                      && 
$129.3$    && 130.1  &&   76.4      & \cr
&                                                                      && 
$139.2$    && 140.2  &&   54.3      & \cr
&                                                                      && 
$149.1$    && 150.0  &&   39.0     & \cr
&                                                                      && 
$159.1$    && 160.0  &&   28.5       & \cr
\colrule
& $pp \to W^\pm Z^0 \to \mu^\pm\nu_\mu\tau^+\tau^-$  &&  && 0.2  & \cr
& $pp \to W^+ W^- \to \mu^+\nu_\mu\tau^-\bar{\nu}_\tau$                &&      
     &&   1.67   & \cr
& $pp \to t\bar{t} \to \mu^\pm\nu_\mu b\tau^\pm\nu_\tau\bar{b}$ &&   && 
$1.37\,10^1$ 
& \cr
& $pp \to Z^0(\gamma^*) \to \tau^+\tau^-\to \mu^+\nu_\mu\bar{\nu}_\tau\tau^-$  
&&           &&   $1.39\,10^4$   & \cr
& $pp \to W^\pm+\mbox{jets} \to \mu^\pm\nu_\mu+\mbox{jets}$ &&  && 
$1.75\,10^{4}$ & \cr
\colrule}}
\end{center}
\end{table*}

\subsection{Hadronic \bm{$\tau$} decay}
\label{sec:hadronic}

The event selection for the hadronic final state of the $\tau$ lepton is 
carried as described below: 
\begin{description}
\item[(1)] Search for one isolated muon ($p_T^\mu > 20$~GeV, $|\eta^\mu| < 
2.5$) to provide the experimental trigger, and one hadronic $\tau$ jet 
($p_T^\tau > 20$~GeV, $|\eta^\tau| < 2.5$). We further require a jet veto 
and a b-jet veto --- no other jet with $p_T > 20$~GeV within $|\eta| < 2.5$ 
--- to reduce $W^\pm +$jets and $t\bar{t} \to \bar{b}\mu^+\nu_\mu 
b\tau^-\bar{\nu}_\tau$ 
backgrounds. A $\tau$ jet identification efficiency of 30\% is assumed.
\item[(2)] The 4-momentum of the $\tau$ lepton is reconstructed from the 
$\tau$ jet and the missing transverse momentum (using the prescription 
of~\cite{HAN}) as 
follows: 
\begin{eqnarray}
\label{eq:pt_tau}
\vec{p}_T^{\;\tau} & = & \vec{p}_T^{\;\tau\mbox{-jet}} + 
\vec{p}_T^{\;\mbox{miss}}, \\
p_z^\tau & = & p_z^{\tau\mbox{-jet}}\, \left(1+ 
\frac{p_T^{\mbox{miss}}}{p_T^{\tau\mbox{-jet}}}\right), \nonumber \\
E^2_\tau & = & \vec{p}^{\; 2}_\tau + m^2_\tau. \nonumber
\end{eqnarray}
The reconstructed momenta of the $\tau$ lepton using 
Equations~(\ref{eq:pt_tau}) are 
shown in Figure~\ref{fig:ptt} together with the generated momenta. 
\begin{figure}
\epsfxsize=8truecm
\begin{center}
     \epsffile{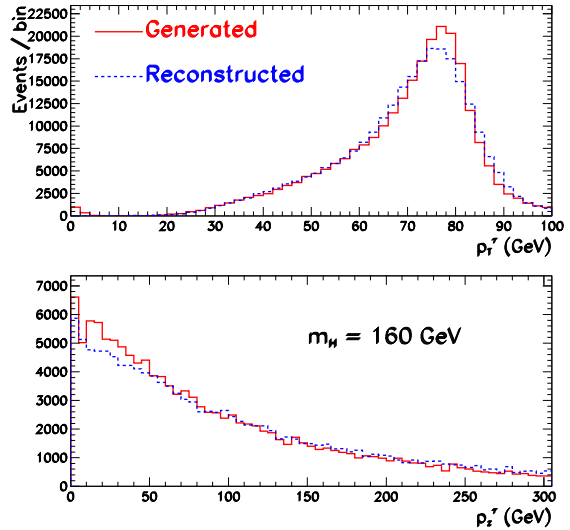}
\caption{The reconstructed and the generated $p_T$ (top plot) and $p_z$ 
(bottom plot) of the $\tau$ lepton. Equations~(\ref{eq:pt_tau}) are used for 
the 
reconstructed 
quantities.}
\label{fig:ptt}
\end{center}
\end{figure}
We demand that the hadronic $\tau$ jet carries at least 60\% of the $\tau$ 
lepton 
energy and the cone $\Delta R = \sqrt{\Delta\eta^2+\Delta\phi^2}$ between the 
$\tau$ jet axis and the $\tau$ lepton direction be less than 0.2 rad:
\begin{eqnarray}
\frac{p_T^{\tau-\mbox{jet}}}{p_T^\tau} & > & 0.6,  \\
\Delta R (p_T^{\tau-\mbox{jet}},p_T^\tau) & < & 0.2\,\mbox{rad}. \nonumber 
\end{eqnarray}  
This cut reduces the background from $W^\pm +\mbox{jets}$ events by more than 
one order 
of magnitude while it costs only a modest $\sim\,40$\% rejection of signal 
events.
\item[(3)] Using the tracker information in the off-line $\tau$ 
identification, we 
require that the $\tau$ jet candidate contains a single charged track within 
$\Delta R < 0.3$~rad around the jet axis. This cut would select one prong 
hadronic 
$\tau$ decay events, and as shown in Figure~\ref{fig:ctracks}, it reduces the 
$W^\pm +\mbox{jets}$ events by an additional factor of ten while costing only 
$\sim$50\% 
reduction in the signal reconstruction efficiencies. 
\begin{figure}
\epsfxsize=8truecm
\begin{center}
\epsffile{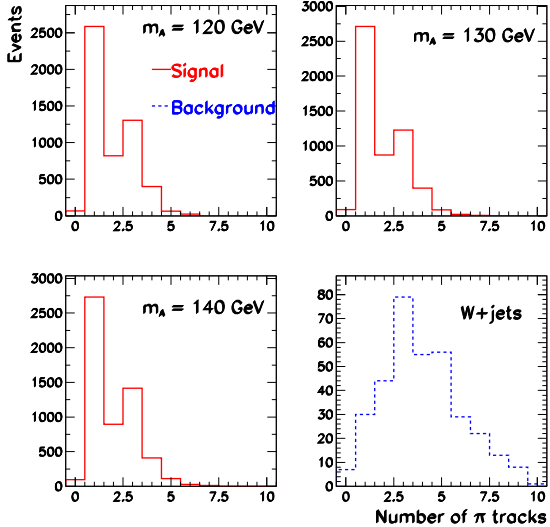}
\caption{The number of reconstructed charged tracks (arbitrary normalization) 
within $\Delta R < 0.3$~rad of the calorimeter jet axis. By requiring a single 
reconstructed charged track so as to select one prong $\tau$ decays, the 
$W^\pm +$jets background is further reduced by one order of magnitude while 
the signal suffers $\sim$ factor of two reduction consistent with the one 
prong hadronic $\tau$ decay branching fraction.}
\label{fig:ctracks}
\end{center}
\end{figure}
\item[(4)] The $\tau$ lepton from the signal is ultra-relativistic, and as a 
result, the missing momentum from $\tau \to (\tau\,\mbox{jet})\nu$ is 
collinear 
with the $\tau$ jet. Further, as a consequence of the two-body decay, the 
$\tau$ jet and the $\mu$ track are back-to-back. We therefore require a large 
azimuthal opening angle between the $\mu$ and the $\tau$ jet and a small 
opening angle between $p_T^{\mbox{miss}}$ and the $\tau$ jet: 
\begin{eqnarray}
\label{eq:opening}
\delta\phi\left(p_T^\mu,p_T^{\tau\mbox{-jet}}\right) & > & 2.75~\mbox{rad}, \\ 
\delta\phi\left(p_T^{\mbox{miss}},p_T^{\tau\mbox{-jet}}\right) & < & 
0.6~\mbox{rad}. 
\nonumber 
\end{eqnarray}
As can be seen from Figure~\ref{fig:htaumu_3}, this cut reduces the signal by 
$\sim$ 35\% while the $pp(\bar{p}) \to Z^0(\gamma^*)$ background is further 
suppressed 
by $\sim$51\%.
\begin{figure}
\epsfxsize=8truecm
\begin{center}
\epsffile{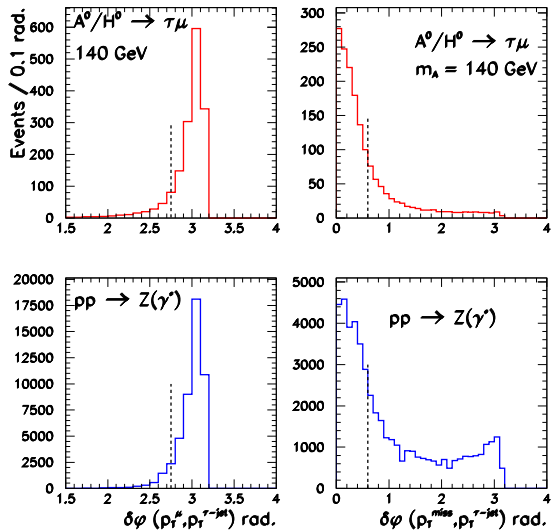}
\caption{The azimuthal opening angle $\delta\phi$ between the muon track and 
the $\tau$ jet (left plots), and between the $p_T^{\mbox{miss}}$ vector and 
the $\tau$ jet (right plots). The $pp(\bar{p}) \to Z^0(\gamma^*)$ background 
is further 
reduced by a factor of two while the signal suffers only a 35\% reduction. The
dashed lines indicate the level of the cuts.}
\label{fig:htaumu_3}
\end{center}
\end{figure}
\item[(5)] The $\mu$ track is mono-energetic because of the two-body decay 
$H^0 \to \tau^\pm\ \mu^\mp$ but the $\tau$ jet in $\tau \to 
(\tau\,\mbox{jet})\nu$ would be somewhat softer. As a result, one would expect 
the momentum difference
\begin{equation}
\label{eq:imbalance}
\Delta p_T = p_T^\mu - p_T^{\tau\mbox{-jet}}
\end{equation}
to be positive for the signal. Indeed, as noted in~\cite{HAN} and as shown in 
Figure~\ref{fig:htaumu_4}, this quantity is very powerful in suppressing the 
$pp(\bar{p}) \to Z^0(\gamma^*)$ background further.
\begin{figure}
\epsfxsize=8truecm
\begin{center}
     \epsffile{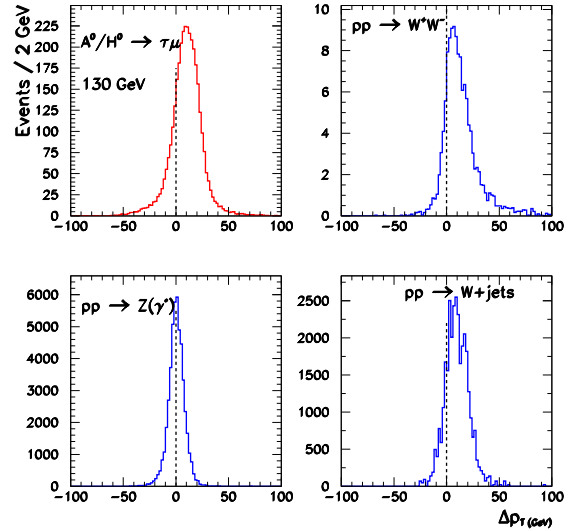}
\caption{The momentum imbalance $\Delta p_T$ between the muon track and the 
$\tau$ jet. In the signal, this quantity is expected to be positive as a 
result of the two-body kinematics from $A^0/H^0 \to \tau^\pm\mu^\mp$ and the 
subsequent decay $\tau \to (\tau\,\mbox{jet})\nu$. This is indeed mostly the 
case as shown in the top left plot. Therefore, demanding $\Delta p_T > 0$ 
suppresses the backgrounds further, particularly the Drell-Yan type process 
$pp(\bar{p}) \to Z^0(\gamma^*)$ which is reduced by as much as 50\% with this 
cut alone as can be seen from the bottom left plot.}
\label{fig:htaumu_4}
\end{center}
\end{figure}
\item[(6)] We now cut on the transverse momentum of the $\tau$ reconstructed 
according to Equations~(\ref{eq:pt_tau}). The distribution of this variable is 
shown in Figure~\ref{fig:pttau} where one sees that demanding $p_T^\tau > 
50$~GeV leads to at most 20\% reduction in the signal --- the $\tau$ gets 
harder at higher $m_A$ so the reduction in the signal due to this cut is 
highest at the lowest mass considered, i.e., $m_A=120$~GeV --- while the 
$W^\pm +$jets and $Z^0(\gamma^*) \to \tau^+\tau^-$ backgrounds are 
suppressed by additional factors of two and ten respectively.
\begin{figure}
\epsfxsize=8truecm
\begin{center}
     \epsffile{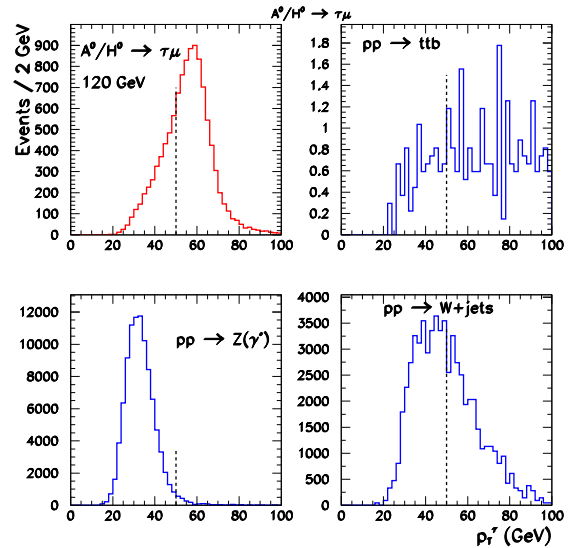}
\caption{The reconstructed transverse momentum of the $\tau$ lepton. We 
require that this quantity be greater 50~GeV leading to additional suppression 
factors of two and ten in the dominant $W^\pm +$jets and $Z^0(\gamma^*) \to 
\tau^+\tau^-$ backgrounds whereas the signal is reduced by at most 20\%.}
\label{fig:pttau}
\end{center}
\end{figure}
\item[(7)] The effective transverse mass of the $\tau\mu$ system 
\begin{equation}
\label{eq:transM}
m_T = \sqrt{2p_T^\mu p_T^{\tau\mbox{jet}}[(1-\cos\delta\phi)]}
\end{equation}
is reconstructed. In the signal, one would expect this quantity to peak toward 
the Higgs mass whereas in the backgrounds, because the final state may 
contain several neutrinos, the $m_T$ distribution would peak at low values as 
shown in Figure~\ref{fig:mt}. 
\begin{figure}
\epsfxsize=8truecm
\begin{center}
     \epsffile{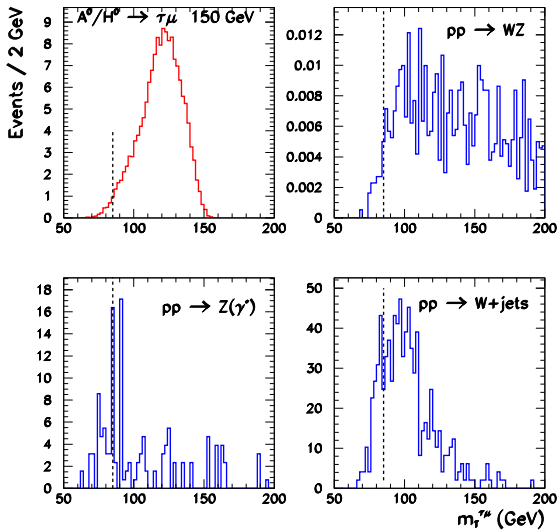}
\caption{The reconstructed effective transverse mass of the $\tau\mu$ system. 
This distribution peaks at low values in the backgrounds while in signal the 
peak is closer to the actual Higgs mass. The dashed lines indicate the cut 
applied on this quantity.}
\label{fig:mt}
\end{center}
\end{figure}
We required that $m_T > 85$~GeV. This cut suppresses the $Z^0(\gamma^*) \to 
\tau^+\tau^-$
background more than the other backgrounds. 
\end{description}
The efficiencies of the cuts discussed above are shown in 
Table~\ref{tab:cut_eff} where one sees that 
the analysis steps described here is effective in reducing the two main 
backgrounds namely $W^\pm +$jets and $Z^0 \to \tau^+\tau^-$. The most 
effective cuts are the ones imposed for the identification of the $\tau$ 
lepton --- cuts \textbf{(2)} and \textbf{(3)} --- and some kinematic cuts such 
as the momentum imbalance defined in cut~\textbf{(5)}. 

\begin{figure}
\epsfxsize=8truecm
\begin{center}
     \epsffile{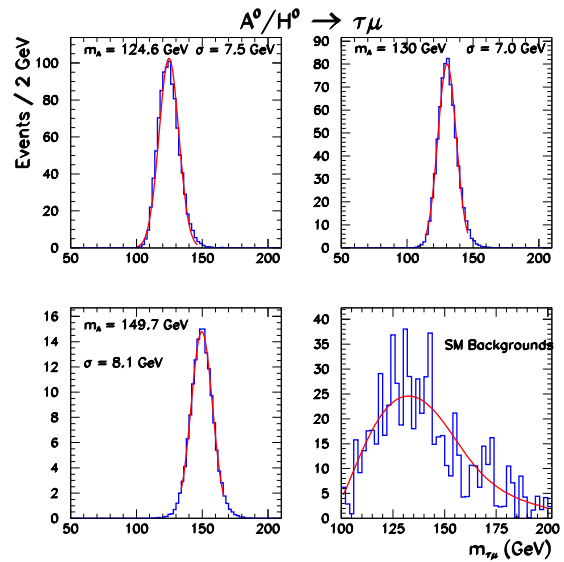}
\caption{The invariant mass $m_{\tau\mu}$ distributions of the signal $A^0/H^0 
\to 
\tau^\pm\mu^\mp$ for several values 
of the Higgs boson mass and also of the backgrounds, after an integrated 
luminosity of 30 fb$^{-1}$. The LFV coupling parameter $\kappa^2_{\tau\mu}=1$. 
The 
signal is reconstructed to within 5~GeV of the Higgs mass above the residual 
$W^\pm +$jets continuum.}
\label{fig:htaumu_5}
\end{center}
\end{figure}

With the $\tau$ 4-momentum $p^\tau$ obtained in Equations~(\ref{eq:pt_tau}), 
the invariant mass of the Higgs boson is reconstructed, 
\begin{equation}
\label{eq:inv}
m^2_{\tau\mu} = (p^\tau + p^\mu)^2. 
\end{equation}
Distributions of $m_{\tau\mu}$ are shown in Figure~\ref{fig:htaumu_5} for the 
signal and the backgrounds.
\begin{table*}
\begin{center}
\begin{minipage}{0.9\linewidth} 
\caption{\label{tab:cut_eff}The efficiencies (in percent) of the cuts used in 
the current analysis. The first three cuts are effective in reducing the 
dominant 
$W^\pm +$jets events while the other cuts suppress the rest of the backgrounds 
efficiently.}
\end{minipage}
\vbox{\offinterlineskip 
\halign{&#& \strut\quad#\hfil\quad\cr  
\colrule
& Cut &  & &  $A^0/H^0$ & $\rightarrow$ & $\tau^\pm\mu^\mp$ & & & &  & \cr 
\colrule
 & & & 120 & 130 & 140 & 150 & 160 GeV & & & $t\bar{t}$ & $W^\pm Z^0$ & 
$W^+W^-$  & $Z^0(\gamma^*)$ & $W^\pm +$jets & \cr \colrule
& \textbf{(1)} &  & 16.3 & 17.0 & 16.9 & 17.2 & 20.9 & & & 3.0  10$^{-1}$ & 
7.2 & 16.3 & 0.2 & 0.5 & \cr
& \textbf{(2)} &  & 9.1  & 9.6  & 9.7  & 9.9  & 12.1 & & & 1.5  10$^{-2}$ & 
0.50 & 1.3 & 0.09 & 2.2 10$^{-2}$ & \cr 
& \textbf{(3)} &  & 4.4  & 4.8  & 4.6  & 4.9  & 5.9 & & & 0.7 10$^{-2}$ & 0.25 
& 0.65 & 0.04 & 2.2 10$^{-2}$ & \cr 
& \textbf{(4)} &  & 2.7  & 3.0  & 2.9  & 3.1  & 3.8 & & & 2.7 10$^{-3}$ & 0.10 
& 0.40 & 0.02 & 1.9 10$^{-3}$ & \cr 
& \textbf{(5)} &  & 2.4  & 2.7  & 2.6  & 2.8  & 3.5 & & & 2.4 10$^{-3}$ & 0.10 
& 0.40 & 0.01 & 1.1 10$^{-3}$ & \cr 
& \textbf{(6)} &  & 1.7  & 2.1  & 2.2  & 2.5  & 3.2 & & & 2.2 10$^{-3}$ & 0.05 
& 0.20 & 2.4 10$^{-4}$ & 5.3 10$^{-4}$ & \cr 
& \textbf{(7)} &  & 1.2  & 2.0  & 2.2  & 2.5  & 3.1 & & & 2.1 10$^{-3}$ & 0.05 
& 0.20 & 1.6 10$^{-4}$ & 4.3 10$^{-4}$ & \cr 
\colrule}}
\end{center}
\end{table*}
We see in this figure that the signal is reconstructed within one GeV of 
the expected Higgs boson 
mass ---  except at $m_A=120$~GeV where the $A^0$ and the $H^0$ are not 
degenerate 
in mass and their summed signal peaks somewhere in the middle  --- while in 
the 
backgrounds the $m_{\tau\mu}$ distribution gives a continuum spectrum 
dominated by 
$W^\pm +$jets event.
\subsection{\bm{$A^0/H^0 \to \tau^\pm\mu^\mp$} versus \bm{$A^0/H^0 \to 
\tau^+\tau^-$}}
\label{sec:A/H}

The $H^0 \to \tau^+\tau^-$ of the SM is not expected to yield a significant 
signal at the LHC due to a low signal rate and substantial backgrounds from 
various sources~\cite{Rainwater}. In the MSSM, for a Higgs boson of the same 
mass, the $A^0/H^0 \to \tau^+\tau^-$ rates are significantly larger than the 
SM case. The $A^0/H^0 \to \tau^+\tau^-$ process has been studied extensively 
for the LHC, and it is demonstrated that such a signal can be observed with a 
significance exceeding $5\sigma$ in a large area of the ($m_A$, $\tan\beta$) 
plane~\cite{CAVA,RITVA}. 

The final state of both processes $A^0/H^0 \to \tau^+\tau^-$ and $A^0/H^0 \to 
\tau^\pm\mu^\mp$ are very similar, namely an isolated $\mu$, a hadronic $\tau$ 
jet and missing energy. The observation of these signals would rely on two 
crucial detector performance parameters, namely a very good 
$p_T^{\mbox{miss}}$ resolution and a very good $\tau$ jet identification with 
excellent rejection of non $\tau$ jets. The former 
performance parameter is necessary for the reconstruction of the $\tau\mu$ 
invariant mass in $A^0/H^0 \to \tau^\pm\mu^\mp$ (as demonstrated in the above 
analysis) and also for the $\tau\tau$ invariant mass in $A^0/H^0 \to 
\tau^+\tau^-$~\cite{CAVA,RITVA} while the latter performance 
parameter allows for the suppression of various backgrounds containing fake 
$\tau$ jets. We show in this section that the reconstruction procedures 
presented in this paper for $A^0/H^0 \to \tau\mu$ and described 
in~\cite{CAVA,RITVA} for $A^0/H^0 \to \tau\tau$ allow for the identification 
of
each of these processes although their final states are similar. 

\subsubsection{Optimization for $A^0/H^0 \to \tau\mu$}
\label{sec:opt_taumu}

\begin{figure}
\epsfxsize=8truecm
\begin{center}
     \epsffile{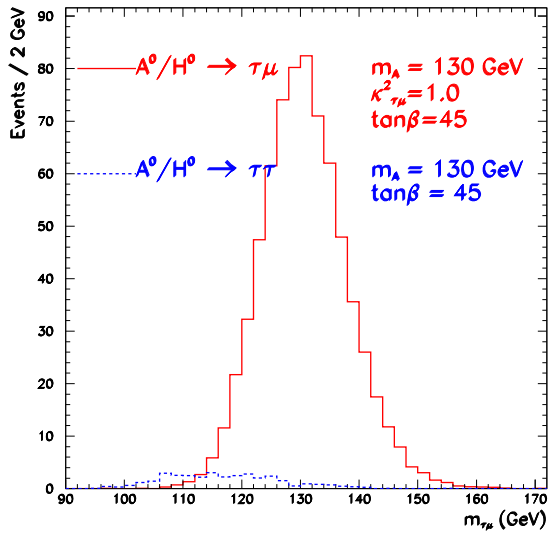}
\caption{The reconstructed $m_{\tau\mu}$ invariant mass for 
$A^0/H^0\to\tau\mu$ and 
$A^0/H^0 \to \tau^+\tau^-$ $m_A=130$~GeV and $\kappa_{\tau\mu}=1$ ($\tan\beta=45$, 
$\alpha=-0.58$~rad), i.e. set~2 of Table~\ref{tab:set} using the analysis procedure presented 
above. The existence of the $A^0/H^0 \to \tau^+\tau^-$ signal would constitute an additional 
background for the $A^0/H^0 \to \tau^\pm\mu^\mp$ process.}
\label{fig:compare1}
\end{center}
\end{figure}

We generated $A^0/H^0 \to \tau\tau$ events and analyzed them according the 
analysis procedure described in section~\ref{sec:hadronic}. The relative 
efficiencies of the cuts described in section~\ref{sec:hadronic} for 
$A^0/H^0\to\tau\mu$ 
and $A^0/H^0\to\tau\tau$ final states are shown in Table~\ref{tab:comp1}. 
\begin{table*}
\begin{center}
\begin{minipage}{0.6\linewidth} 
\caption{\label{tab:comp1}The relative efficiencies (in percent) of the cuts 
used in the current analysis for $A^0/H^0\to\tau\tau$ events, to be compared to 
Table~\ref{tab:cut_eff} where the efficiencies for $A^0/H^0\to\tau\mu$ are shown.}
\end{minipage}
\vbox{\offinterlineskip 
\halign{&#& \strut\quad#\hfil\quad\cr  
\colrule
& Cut & & & $A^0/H^0$ & $\to$ 
& $\tau\tau$  & \cr 
\colrule
&               & & 120   & 130   & 140 & 150   & 160 GeV & \cr 
\colrule
& \textbf{(1)} & &  2.0  & 4.2  & 2.1  & 1.95  & 1.77     & \cr
& \textbf{(2)} & &  0.6  & 1.1  & 0.6  & 0.52  & 0.47     & \cr 
& \textbf{(3)} & &  0.3  & 0.6  & 0.3  & 0.26  & 0.23     & \cr 
& \textbf{(4)} & &  0.1  & 0.24 & 0.12  &  0.11  &  0.10    & \cr 
& \textbf{(5)} & &  0.09  & 0.18 &  0.09 &  0.08  &  0.08     & \cr 
& \textbf{(6)} & &  0.01 &  0.04  &  0.02 &  0.03 &  0.03     & \cr 
& \textbf{(7)} & &  0.8$\,10^{-2}$ &  0.02 &  0.02 &  0.02 &  0.03    & \cr 
\colrule}}
\end{center}
\end{table*}
From Figure~\ref{fig:compare1}, we see that at the same Higgs boson mass, the 
reconstructed $\tau\mu$ invariant mass for the $A^0/H^0 \to \tau^+\tau^-$ 
events peaks at lower values. 

\subsubsection{Optimization for $A^0/H^0 \to \tau\tau$}
\label{sec:opt_tautau}

It is also important to show that the analysis technique optimized for the 
search for the $A^0/H^0\to\tau\tau$ signal is capable of separating the 
$\tau\tau$ final state from the $\tau\mu$ events. We have therefore examined 
$A^0/H^0\to\tau\mu$ events according to the $A^0/H^0\to\tau\tau$ analysis 
technique which we recall succinctly as follows~\cite{CAVA,RITVA}:
\begin{table*}
\begin{center}
\begin{minipage}{0.6\linewidth} 
\caption{\label{tab:comp2}The relative efficiencies (in percent) of the cuts 
used in the search for $A^0/H^0\to\tau\tau$~\cite{CAVA,RITVA} and re-stated 
briefly in the text ($m_A=130$~GeV).}
\end{minipage}
\vbox{\offinterlineskip 
\halign{&#& \strut\quad#\hfil\quad\cr  
\colrule
& Cut &  & $A^0/H^0 \to\tau^\pm\mu^\mp$ & &  & $A^0/H^0\to\tau\tau$  & \cr 
\colrule
& \textbf{(a)} &    & 15.1               & &  & 3.1  & \cr
& \textbf{(b)} &    & 5.3               & &  & 1.9   & \cr 
& \textbf{(c)} &    & 0.2               & &  & 1.5  & \cr 
& \textbf{(d)} &    & 0.02               & &  & 0.3  & \cr 
\colrule}}
\end{center}
\end{table*}
\begin{description}
\item[(a)] One isolated $\mu$ with $p_T > 24$~GeV and $|\eta| < 2.5$, one 
hadronic $\tau$ jet with $E_T^{jet} > 40$ and $|\eta| > 2.5$ and b-jet veto.
\item[(b)] $E_T^{miss} > 18$~GeV.
\item[(c)] The transverse mass $m_T$(lepton-$E_T^{miss}$)$< 25$~GeV.
\item[(d)] $1.8 < \Delta\phi < 2.9$~rad or $3.4 < \Delta\phi < 4.9$~rad, where 
$\Delta\phi$ is the azimuthal opening angle between the $\tau$ jet and the 
isolated $\mu$. This cut is needed for the reconstruction of the $\tau\tau$ 
invariant mass $m_{\tau\tau}$. Indeed, the invariant mass $m_{\tau\tau}$ 
of the pair of $\tau$ leptons produced in the process
\[ A^0/H^0 \to \tau\tau \to \mbox{jet}\,\nu_\tau\mu\nu_\mu\nu_\tau \]
can be reconstructed assuming that $m_\tau=0$, that the $\tau$ detected 
products 
(in this case the $\tau$ jet and the $\mu$) are not back-to-back, and also 
that the 
direction of the neutrino system from each $\tau$ decay coincides with 
that of 
the detected product:
\begin{equation}
\label{eq:mtt}
m_{\tau\tau} = \sqrt{2(E_1+E_{\nu 1})(E_2+E_{\nu 2})(1-\cos\theta)}.
\end{equation}
$E_1$ and $E_2$ are the visible energies from the $\tau$ decays, $\theta$ is 
the angle 
between the directions of the detected products, and $E_{\nu 1}$ and $E_{\nu 
2}$ are the 
energies of the two neutrino systems, obtained by solving the system of 
equations
\[ p_x^{miss}\left(p_y^{miss}\right) = \left[E_{\nu 1}\bar{u}_1\right]_{x(y)}+
\left[E_{\nu 2}\bar{u}_2\right]_{x(y)}, \]
where $\bar{u}_1$ and $\bar{u}_1$ are the directions of the detected products, 
and 
$p_x^{miss}$ and $p_y^{miss}$ the components of the $E_T^{miss}$ vector. The 
above 
system of equations can be solved if the determinant, which is proportional to 
$\sin\Delta\phi$, is not zero. Further details of the $m_{\tau\tau}$ 
reconstruction 
are well documented elsewhere~\cite{CAVA,RITVA}. 
\end{description}
\begin{figure}
\epsfxsize=8truecm
\begin{center}
     \epsffile{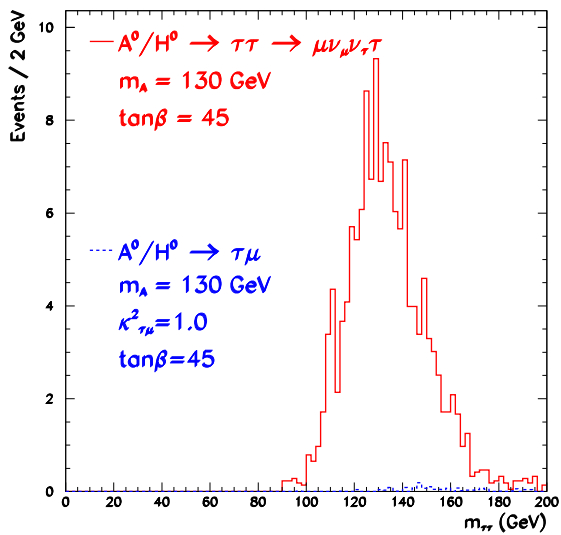}
\caption{The reconstructed $m_{\tau\tau}$ invariant mass for 
$A^0/H^0\to\tau\tau$ and $A^0/H^0 \to \tau\mu$, for $m_A=130$~GeV and $\kappa_{\tau\mu}=1$ 
($\tan\beta=45$, $\alpha=-0.58$~rad), i.e. set~2 of Table~\ref{tab:set} using the 
analysis procedure presented in~\cite{CAVA,RITVA}. The existence of the $A^0/H^0 \to \tau\mu$ 
signal would constitute a negligible background for the $A^0/H^0 \to \tau\tau$ process.}
\label{fig:compare2}
\end{center}
\end{figure}
The relative efficiencies of the cuts \textbf{(a)}--\textbf{(d)} for the 
$\tau\mu$ and $\tau\tau$ final states are shown in Table~\ref{tab:comp2}. 
Figure~\ref{fig:compare2} shows the reconstructed $m_{\tau\tau}$ invariant 
mass distribution for both final states. The $\tau\mu$ events would contribute 
a negligible background under the $\tau\tau$ signal.

The reconstruction procedure for $A^0/H^0 \to \tau^+\tau^-$ 
described in~\cite{CAVA,RITVA} and the analysis steps presented above for 
$A^0/H^0 \to \tau^\pm\mu^\mp$, would allow for the separation of 
both signals, with each contributing a small residual background under peak of 
the other as shown in Figures~\ref{fig:compare1} and~\ref{fig:compare2}. 

\subsection{Leptonic \bm{$\tau$} decay}
\label{sec:lep_decay}

Thus far, we have considered the hadronic final state of the $\tau$ lepton, 
and the major 
irreducible background comes from $W+$jets events where a jet is 
mis-identified as a hadronic 
$\tau$ jet. Indeed, the residual SM background shown in 
Figure~\ref{fig:htaumu_5} is dominated by 
$W+$jets events whose rate is several orders of magnitude higher than the 
signal rates as shown in 
Table~\ref{tab:htaumu_1}. In this section, we examine the leptonic decay of 
the $\tau$, namely 
$\tau \to e\nu_e\bar{\nu}_\tau$. Although the branching fraction of 
$\tau \to e\nu_e\bar{\nu}_\tau$ is only 
$\sim$ 18\% compared to 65\% for $\tau \to (jet)\nu_\tau$, the identification 
of the electron is 
easier with an efficiency of 90\% whereas the $\tau$ jet identification 
efficiency is much lower: 
in the above analysis, we assume a $\tau$ jet identification efficiency of 
30\%, corresponding to 
a jet rejection factor of $\sim$ 400 --- see~\cite{CAVA} for details. 
\begin{figure}
\epsfxsize=8truecm
\begin{center}
\epsffile{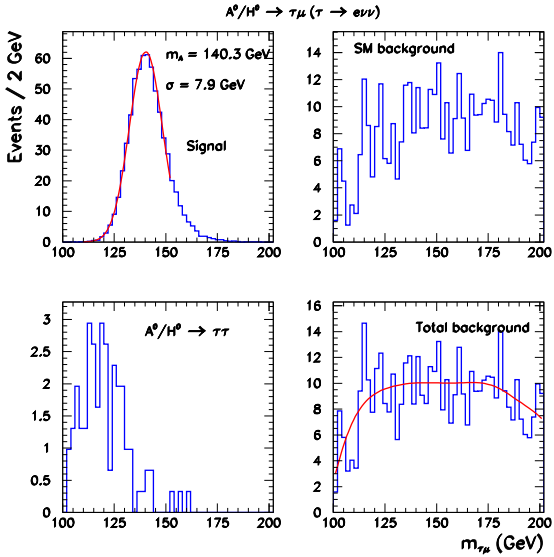}
\caption{The reconstructed invariant mass $m_{\tau\mu}$ in the leptonic decay 
of the 
$\tau$ ($\to e\nu\nu$) for the signal ($m_A=140$~GeV, $\tan\beta=45$, 
$\kappa=1$), the 
SM backgrounds and the $A^0/H^0 \to \tau\tau$ background.}
\label{fig:e140}
\end{center}
\end{figure}
\begin{table*}
\begin{center}
\begin{minipage}{.75\linewidth} 
\caption{\label{tab:TeV} The rates, $\sigma\,\times\,$~BR(pb), for the 
signal $gg\to A^0/H^0 \to \tau^+\mu^- + \tau^-\mu^+$, and the  backgrounds at 
the TeVatron. We assume $\kappa_{\tau\mu} = 1$ and $\tan\beta=45$.}
\end{minipage}
\vbox{\offinterlineskip 
\halign{&#& \strut\quad#\hfil\quad\cr  
\colrule
& Process   && $m_A$ (GeV)  && $m_H$ (GeV) && $\sigma\,\times\,$~BR(pb)   & \cr
\colrule
& $gg\to A^0/H^0 \to \tau^+\mu^- + \tau^-\mu^+$                            && 
$119.3$  && 128.4    &&   $1.41\, 10^{-1}$     & \cr
&                                                                      && 
$129.3$  && 130.1    &&   $0.79\, 10^{-1}$      & \cr
&                                                                      && 
$139.2$  && 140.2    &&   $0.33\, 10^{-1}$      & \cr
&                                                                      && 
$149.1$  && 150.0     &&   $0.11\, 10^{-1}$      & \cr
&                                                                      && 
$159.1$  && 160.0     &&   $0.15\, 10^{-2}$       & \cr
\colrule
\colrule
& $gg\to A^0/H^0 \to \tau\tau$                                         && 
$119.3$    && 128.4  &&   3.90     & \cr
&                                                                      && 
$129.3$    && 130.1  &&   2.84      & \cr
&                                                                      && 
$139.2$    && 140.2  &&   1.79      & \cr
&                                                                      && 
$149.1$    && 150.0  &&   1.16     & \cr
&                                                                      && 
$159.1$    && 160.0  &&   0.75       & \cr
\colrule
& $p\bar{p} \to Z^0(\gamma^*) \to \tau^+\tau^-\to 
\mu^+\nu_\mu\bar{\nu}_\tau\tau^-$  
&&           &&   $3.24\,10^3$   & \cr
& $p\bar{p} \to W^\pm+\mbox{jets} \to \mu^\pm\nu_\mu+\mbox{jets}$ &&  && 
$3.21\,10^{3}$ & \cr
\colrule}}
\end{center}
\end{table*}
Furthermore, the leptonic 
decay of the $\tau$ will not be sensitive to the $W+$jets background. We 
search for a signal final state containing two isolated leptons, one electron 
and the other 
a $\mu$ with no hadronic activity. The major SM backgrounds in this case are 
shown in 
Table~\ref{tab:htaumu_1} --- processes listed in~(\ref{eq:backg}) --- except for the 
$W+$jets background, in addition to:
\begin{eqnarray}
\label{eq:add_bgd}
t\bar{t} & \to & \mu\nu_\mu b e\nu_e\bar{b}, \\
WW       & \to & \mu\nu_\mu e\nu_e. \nonumber
\end{eqnarray}
\begin{figure}
\epsfxsize=8truecm
\begin{center}
     \epsffile{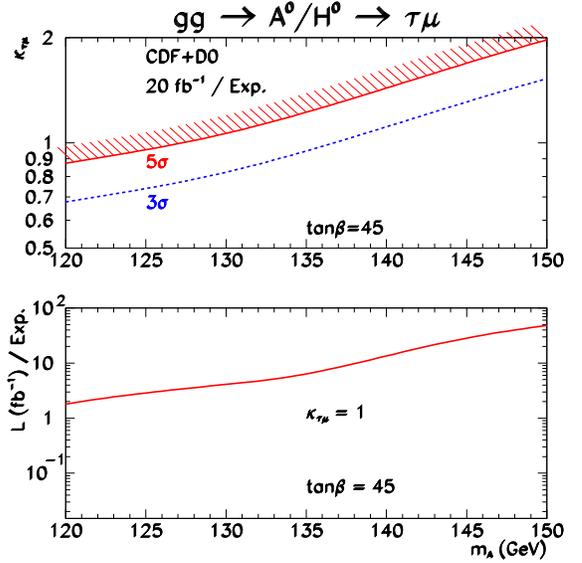}
\caption{The discovery reach at the TeVatron in the combined 
$\tau\to\mbox{jet}\,\nu_\tau$ and $\tau \to e\nu_e\nu_\tau$ channels. The 
signal would yield a 
$5\sigma$ significance for $0.87 \OOle \kappa_{\tau\mu} \OOle 2.0$ and for a 
Higgs boson mass $120 \OOle m_A \OOle 150$~GeV. The luminosity needed for a 
95\% confidence level exclusion is shown in the bottom plot.}
\label{fig:disco_tev}
\end{center}
\end{figure}
 The reconstruction of the signal is 
exactly as described in section~\ref{sec:hadronic}, except for the 
cuts~\textbf{(2)} and~\textbf{(3)} 
which were implemented for the suppression of the $W+$jets events and for the 
selection of the one prong 
hadronic $\tau$ decays. These cuts are no longer necessary and are not used in 
the search for the leptonic decay of the $\tau$. Figure~\ref{fig:e140} shows 
the
 reconstructed $\tau\mu$ invariant mass for the signal $A^0/H^0 \to \tau\mu$, 
the SM 
backgrounds and for the $A^0/H^0 \to \tau\tau$ background with one $\tau$ 
decaying to 
leptons: $\tau\to e\nu\nu$. In this channel too, the signal can be observed 
with
significances exceeding $5\sigma$ depending on the LFV coupling parameter 
$\kappa_{\tau\mu}$. 

\subsection{Prospects at the TeVatron}
\label{sec:prospects_TeV}

Table~\ref{tab:TeV} shows the estimated signal and background rates at the 
TeVatron where we propose to search for $A^0/H^0 \to \tau^\pm\mu^\mp$ with 
the neutral Higgs bosons of the 2HDM produced through gluon fusion: $gg \to 
A^0/H^0$.
The signal-to-background ratios and the signal significances calculated  
within 
$\pm 2\sigma$ of the reconstructed Higgs mass peak, for an integrated 
luminosity of 
20~fb$^{-1}$ per experiment, are shown in Table~\ref{tab:sign_tev} for 
$\kappa^2_{\tau\mu}=1$ and $\tan\beta=45$.
\begin{table*}
\begin{center}
\begin{minipage}{0.65\linewidth} 
\caption{\label{tab:sign_tev}The expected signal-to-background ratios and 
signal 
significances ($\tau \to \mbox{jet}\,\nu$/$\tau \to e\nu\nu$) for two 
experiments 
at the TeVatron, assuming $\kappa^2_{\tau\mu}=1$ and 5\% systematic 
uncertainty on 
the background shape and normalization.}
\end{minipage}
\vbox{\offinterlineskip 
\halign{&#& \strut\quad#\hfil\quad\cr  
\colrule
&  $m_A$ (GeV) $\to$    && 120      && 130      && 140      && 150      & \cr 
\colrule
& Signal ($S$)          && 10/29    && 7/19     && 3/13     &&  1/5     & \cr
& Backgrounds ($B$)     && 4/42     && 4/44     && 3/51     &&  2/62    & \cr
& $S/B$                 && 2.4/0.7  && 1.8/0.4  && 1.0/0.3  &&  0.5/0.1 & \cr
& $S/\sqrt{B}$          && 5.0/4.3  && 3.5/2.7  && 1.7/1.7  &&  1.2/0.6 & \cr
& Combined $S/\sqrt{B}$ && 6.6      && 4.4      && 2.4      &&  1.3     & \cr
\colrule}}
\end{center}
\end{table*}
At the TeVatron, a significant signal ($>5\sigma$) can be detected for Higgs 
boson masses
around 120~GeV and high $\tan\beta$ ($\sim 45$), assuming $\kappa_{\tau\mu} 
\sim 1$. 
\begin{figure}
\epsfxsize=8truecm
\begin{center}
     \epsffile{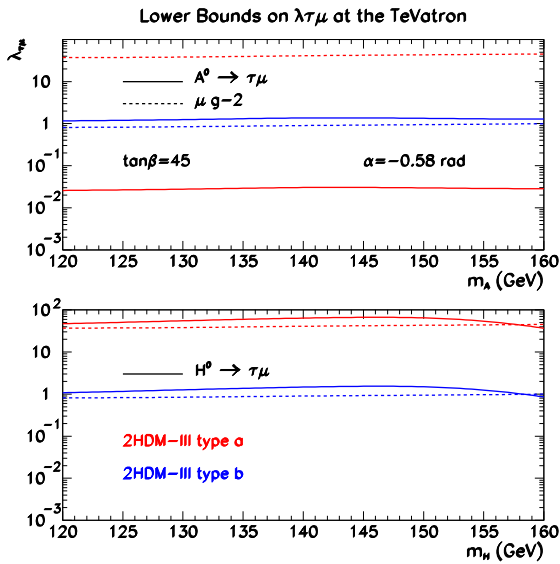}
\caption{The achievable lower bounds on the LHV coupling $\lambda_{\tau\mu}$ 
at the TeVatron (CDF $+$ D\O) obtained from the summed signal in the hadronic 
and leptonic decays of the $\tau$ lepton. To obtain these bounds, we use the 
values of $\kappa_{\tau\mu}$ at $5\sigma$ shown in Figure~\ref{fig:disco_tev} --- 20~fb$^{-1}$ 
per experiment. The current bounds on $\lambda_{\tau\mu}$ obtained from the muon 
$g-2$ data are also shown. For the muon $g-2$ data, the curves at higher $\lambda_{\tau\mu}$ 
values correspond to the 2HDM-III type a and the lower curves to the 2HDM-III type b. 
This trend is the same for $H^0\to\tau\mu$ (bottom plot) but it is reversed for 
$A^0\to\tau\mu$ (top plot).}
\label{fig:tev_lambda}
\end{center}
\end{figure}
We show in Figure~\ref{fig:disco_tev} the discovery reach at the TeVatron and 
the 
luminosity required for a 95\% confidence level exclusion for large 
$\tan\beta$. For 
low $\tan\beta$ values ($\OOle 10$), the signal production rate decreases by 
more than 
an order of magnitude compared to the case shown in Table~\ref{tab:TeV} so 
that the 
detection of this process at the TeVatron would require very large values of 
the LFV 
coupling $\lambda_{\tau\mu}$. 
\begin{figure}
\epsfxsize=8truecm
\begin{center}
\epsffile{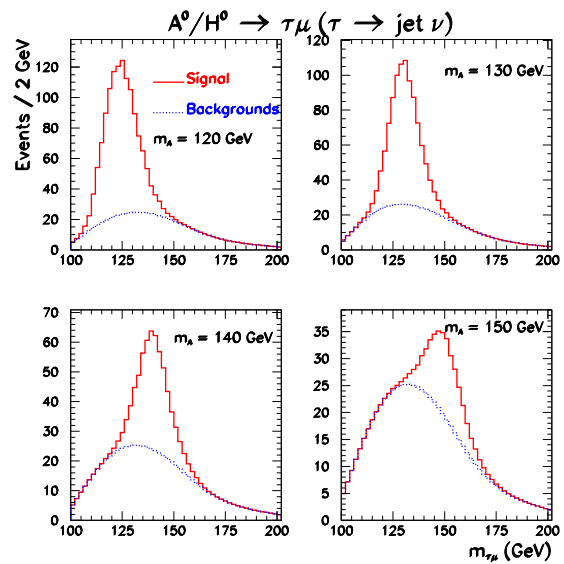}
\caption{The reconstructed invariant mass $m_{\tau\mu}$, after 
cut~\textbf{(7)}, of the signal plus the backgrounds in the hadronic 
$\tau$ decay channel for $m_A=120$, 130, 140 and 150~GeV,  and for an 
integrated luminosity of 30 fb$^{-1}$ at the LHC. For the assumed value of the 
LFV 
coupling parameter ($\kappa^2_{\tau\mu}=1$), the signal can be observed with a 
significance exceeding 5$\sigma$ up to $m_A=150$~GeV.}
\label{fig:inv}
\end{center}
\end{figure}
\begin{figure}
\epsfxsize=8truecm
\begin{center}
\epsffile{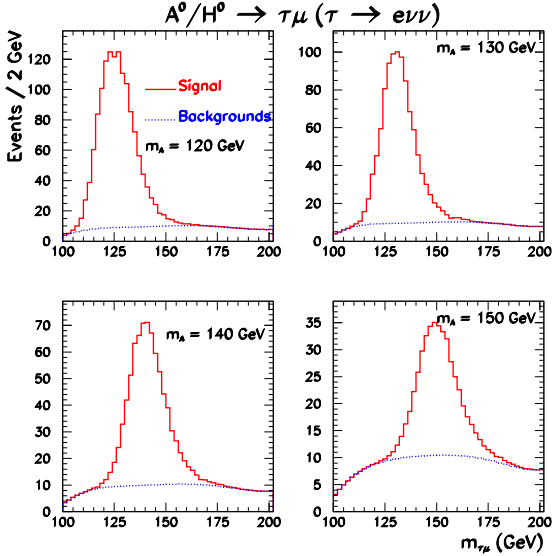}
\caption{The same as Figure~\ref{fig:inv} but with the leptonic decay of the 
$\tau$ ($\to e\nu_e\nu_\tau$).}
\label{fig:inv_e}
\end{center}
\end{figure}
However, one would expect the LFV couplings 
$\lambda_{ij} \sim {\cal O}(1)$~\cite{HAN,SHER2} --- see 
Equations~(\ref{eq:kl1}) 
and~(\ref{eq:kl2}). Therefore, at the TeVatron, this channel would be viable 
only in the 
event of a large $\tan\beta$ value and for $\kappa_{\tau\mu}\sim 1$ --- see 
Table~\ref{tab:k-l} for the correspondence between $\kappa_{\tau\mu}$ and 
$\lambda_{\tau\mu}$. 
Figure~\ref{fig:tev_lambda} shows the corresponding 
expected limits on 
$\lambda_{\tau\mu}$ at the TeVatron: the reach in $\lambda_{\tau\mu}$ would be 
extended, 
at large $\tan\beta$,  beyond that obtained from the muon $g-2$ experiment, for 
$A^0 \to \tau\mu$ in the 2HDM-III type a. 

As shown in Figure~\ref{fig:tev_lambda}, the bounds on $\lambda_{\tau\mu}$ would 
be different for the Higgs bosons $A^0$, $H^0$ and $h^0$ because of their different 
LFV Yukawa couplings --- see sections~\ref{sec:lowen}-\ref{sec:colliders} and the appendix. 

\subsection{Prospects at the LHC}
\label{sec:prospects_LHC}

The signal and background rates at the LHC are shown in 
Table~\ref{tab:htaumu_1}. In 
Figures~\ref{fig:inv}~and~\ref{fig:inv_e}, we show the reconstructed 
$m_{\tau\mu}$ 
invariant mass for several values of the Higgs mass, and for an integrated 
luminosity 
of 30~fb$^{-1}$, and for the LFV coupling parameter $\kappa^2_{\tau\mu}=1$.

\begin{table*}
\begin{center}
\begin{minipage}{0.8\linewidth} 
\caption{\label{tab:sign}The signal-to-background ratios and signal 
significances calculated within $\pm 2\sigma$ of the reconstructed Higgs mass 
$<m_A>$ for $\tau \to \mbox{jet}\,\nu$/$\tau \to e\nu\nu$ --- one experiment 
at the LHC --- with  
an integrated luminosity of 30~fb$^{-1}$, assuming 
$\kappa^2_{\tau\mu}=1$ and 5\% systematic uncertainty from the residual 
background shape and normalization.}
\end{minipage}
\vbox{\offinterlineskip 
\halign{&#& \strut\quad#\hfil\quad\cr  
\colrule
&  $m_A$ (GeV) $\to$    && 120   && 130   && 140   && 150   && 160   & \cr 
\colrule
& $<m_A>$  (GeV)        && 124.6/125.2 && 130.0/130.7 && 139.9/140.6 && 
149.7/150.0 && 159.4/159.8 & \cr
& $\sigma$ (GeV)        &&   7.5/7.3   &&   7.0/7.0   &&   7.6/8.2   &&   
8.1/9.1   &&   8.4/10.4 & \cr
& Signal ($S$)          && 943/1142    && 687/816     && 349/624     &&  
144/279    &&   23/57  & \cr
& Backgrounds ($B$)     && 360/134     && 397/140     && 376/163     &&  
296/198    && 223/226   & \cr
& $S/B$                 && 2.6/8.5     && 1.7/5.8     && 0.9/3.8     &&  
0.5/1.4    &&  0.1/0.3  & \cr
& $S/\sqrt{B}$          && 36.1/85.4   && 24.4/59.4   && 12.9/41.2   &&  
6.3/16.2   && 1.2/3.0   & \cr
& Combined $S/\sqrt{B}$ && 92.7        && 64.2        && 43.2        &&  17.4  
     && 3.2       & \cr
\colrule}}
\end{center}
\end{table*}
The signal-to-background ratios and the signal significances are calculated 
with the events reconstructed within $\pm 2\sigma$ of the reconstructed Higgs 
mass. As shown in Table~\ref{tab:sign}, a significant signal can be observed 
at the LHC for Higgs masses in the range 120 to 150~GeV for the LFV coupling 
parameter $\kappa_{\tau\mu} \sim {\cal O}(1)$. Around 160~GeV, as the 
$H^0_{SM} \to W^+ W^-$ channel opens up, the rate for $A^0/H^0 \to 
\tau^\pm\mu^\mp$ 
decreases so drastically that the observation of a significant signal would be 
possible only in the event of $\kappa_{\tau\mu} > 1$.

\begin{figure}
\epsfxsize=8truecm
\begin{center}
     \epsffile{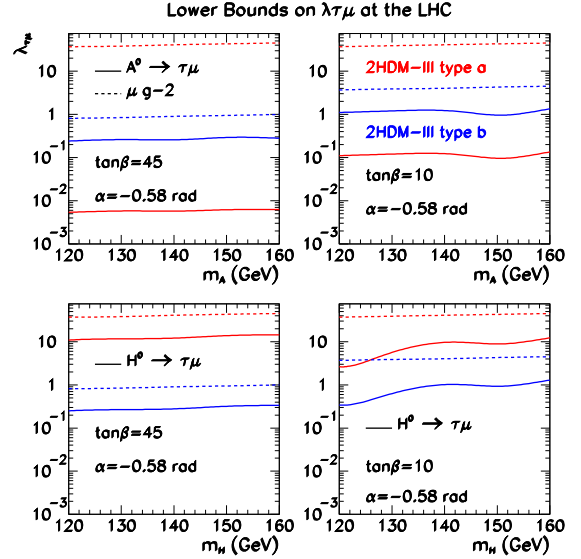}
\caption{The achievable bounds on the LHV coupling $\lambda_{\tau\mu}$ at 
the LHC (ATLAS $+$ CMS) from the combined hadronic and leptonic $\tau$ 
decay channels --- 100~fb$^{-1}$ per experiment. For low or high $\tan\beta$, 
the lower bounds on $\lambda_{\tau\mu}$ at the LHC would be extended beyond that 
of the muon $g-2$ experiment. For the muon $g-2$ data, the curves at higher 
$\lambda_{\tau\mu}$ values correspond to the 2HDM-III type a and the lower 
curves to the 2HDM-III type b. This trend is the same for $H^0\to\tau\mu$ (bottom plots) 
but it is reversed for $A^0\to\tau\mu$ (top plots).}
\label{fig:lhc_lambda}
\end{center}
\end{figure}

The constraints on this LFV coupling $\lambda_{\tau\mu}$ from low energy 
experiments are 
rather weak --- see the discussion in section~\ref{sec:lowen} on low 
energy bounds.  
\begin{figure}
\epsfxsize=8truecm
\begin{center}
     \epsffile{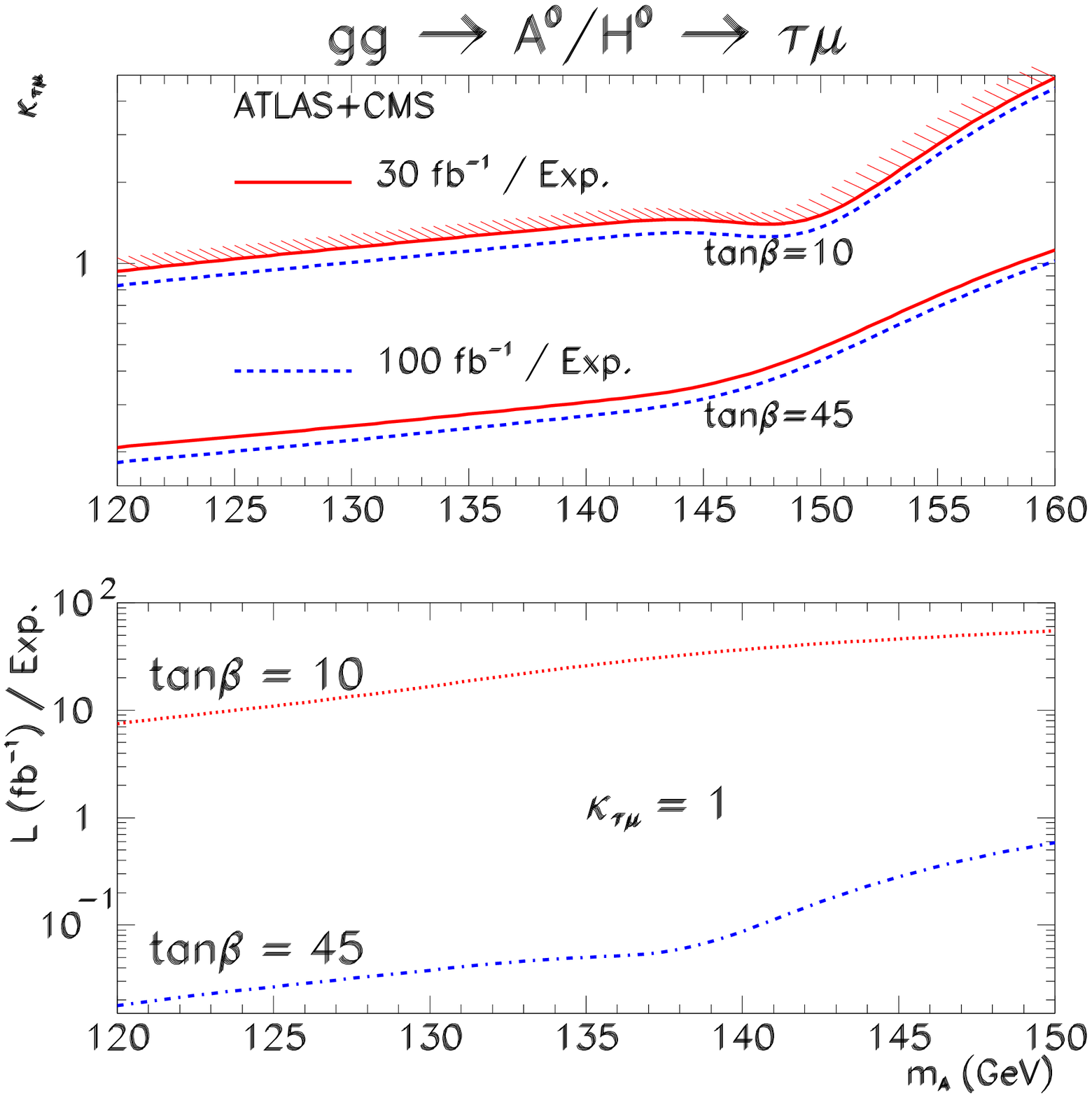}
\caption{The 5$\sigma$ discovery reach in the ($m_A$, $\kappa_{\tau\mu}$) 
plane, using the combined $\tau\to\mbox{jet}\,\nu_\tau$ and $\tau \to 
e\nu_e\nu_\tau$ signals, for ATLAS $+$ CMS (top plot). The $A^0/H^0 \to 
\tau^\pm\mu^\mp$ signal would yield a $5\sigma$ significance for 
$0.18 \OOle \kappa_{\tau\mu} \OOle 1.0$ and for a Higgs boson mass 
$120 \OOle m_A \OOle 160$~GeV. The bottom plot shows the luminosity needed for 
a
95\% confidence level exclusion as a function of $m_A$ for low and high 
$\tan\beta$ 
assuming $\kappa_{\tau\mu}=1$.}
\label{fig:discovery}
\end{center}
\end{figure}
From Equation~(\ref{eq:br-taumu}), the signal rate 
scales like $\kappa^2_{\tau\mu}$ and we show in 
Figure~\ref{fig:discovery} the value of $\kappa_{\tau\mu}$ at which the signal 
yield a $5\sigma$ significance around the Higgs boson mass peak. 
The LFV coupling $0.18 \OOle \kappa_{\tau\mu} \OOle 1.0$ can be reached at 
the LHC, combining ATLAS and CMS data for Higgs boson masses $120 \OOle m_A 
\OOle 
160$~GeV. Figure~\ref{fig:discovery} also shows in the bottom plot, the 
luminosity 
needed at the LHC to achieve a $2\sigma$ (95\% CL) exclusion. At the LHC, 
assuming the 
LFV coupling parameter $\kappa_{\tau\mu} \sim {\cal O}(1)$, few years of low 
luminosity data would be enough to exclude this model in mass 
range $120 < m_A < 150$~GeV and at low $\tan\beta$. For high $\tan\beta$  
values, a 95\% CL exclusion can be established in one year of data taking or 
less for the mass range considered. In Figure~\ref{fig:lhc_lambda}, we show 
the expected bounds on $\lambda_{\tau\mu}$ at the LHC using the values of 
$\kappa_{\tau\mu}$ at $5\sigma$ from 
Figure~\ref{fig:discovery} and we see that the reach in $\lambda_{\tau\mu}$ 
would be extended beyond the muon $g-2$ limits.

\section{Conclusions}
\label{sec:conclusions}

In models with several Higgs doublets, FCNC and LFV couplings exist at tree 
level because the diagonalization of the up-type and the down-type mass 
matrices does not ensure the diagonalization of the Higgs-fermion coupling 
matrices. In the 2HDM-I and II, a discrete symmetry suppresses FCNC and LFV 
couplings at tree level by restricting fermions of a given electric charge to 
couple to at most one Higgs doublet. In the 2HDM-III, no discrete symmetry 
is invoked and the flavor changing couplings are parameterized in terms of 
the fermion mass hierarchy to be in agreement with the severe experimental 
constraints on FCNC and LFV couplings with the first generation index. The 
arbitrariness of the FCNC and LFV couplings of the second
and the third generations can be constrained in low energy and collider 
experiments. The deviation of the measured muon anomalous magnetic moment from 
the SM prediction offers weak bounds on the LFV coupling 
parameter $\lambda_{\tau\mu}$. 

We have investigated the achievable bound on $\kappa_{\tau\mu}$ and 
$\lambda_{\tau\mu}$ at hadron 
colliders by studying the $gg \to A^0/H^0 \to \tau^\pm\mu^\mp$ signal 
observability. 

Considering the hadronic decay of the $\tau$ lepton, the main 
backgrounds of this process are $Z^0(\gamma) \to \tau^+\tau^-$ and $W^\pm 
+$jets 
events where $W^\pm \to \mu^\pm\nu_\mu$ and a jet is mis-identified as a 
$\tau$ jet. 
We search for an isolated $\mu$ and one $\tau$ jet, and we applied a jet veto 
and a 
b-jet veto to reject multi-jet final states from $t\bar{t}$ and 
$W^\pm +$jets. Further reduction of the backgrounds is achieved by exploiting 
the differences in the event topology of the signal and the various 
backgrounds. Although the background rates 
are several orders of magnitude higher than the signal rate, three main 
detector performance parameters have 
been crucial in extracting a significant signal: a good $\tau$ jet 
identification and rejection against non $\tau$ jets, the 
tracking capability for the identification of the charged tracks in one prong 
hadronic $\tau$ decays, and the missing momentum resolution.  

We also investigated the leptonic decay of the $\tau$ ($\to e\nu\nu$) where 
the $W+$jets 
event do not contribute a significant background. In this case, we require a 
final state 
containing one isolated $\mu$, one isolated electron, and we use jet veto and 
b-jet veto 
to suppress the $t\bar{t}$ background. The leptonic decay of the $\tau$ gives 
a better 
sensitivity. The signal significances are estimated based on expected events 
in both the 
hadronic and the leptonic $\tau$ decay channels. 

The analysis steps described above reconstruct the $\tau\mu$ invariant mass to 
within 
$\sim 1$~GeV of the Higgs boson mass (except at $m_A=120$~GeV where the $A^0$ 
and the $H^0$ bosons are not degenerate and the summed signal peaks somewhere 
in the middle as a 
result), and 
also differentiate the $A^0/H^0 \to \tau^+\tau^-$ events from the 
$A^0/H^0 \to \tau^\pm\mu^\mp$ signal.

With an integrated luminosity of 20 fb$^{-1}$ at the TeVatron, a signal for 
$0.87 \OOle \kappa_{\tau\mu} \OOle 2.0$ could be detected with a significance 
of 
$5\sigma$ for Higgs boson masses $120 \OOle m_A \OOle 150$~GeV, corresponding 
to a reach in
$\lambda_{\tau\mu}$ of the order 0.03/1.0 for the 2HDM-III type a/type b. In 
case the signal 
is not observed and assuming $\kappa_{\tau\mu} \sim {\cal O}(1)$, a 95\% CL 
exclusion can be set with $<$ 14~fb$^{-1}$ of data for $120 \OOle m_A \OOle 
140$~GeV. 

The sensitivity will be improved at the LHC where $\kappa_{\tau\mu} \sim 0.18$ 
could be 
reached with 100~fb$^{-1}$ --- this would correspond to a lower bound in 
$\lambda_{\tau\mu}\sim 0.01/0.1$ 
for the 2HDM-III type a/type b. A 95\% CL exclusion could be set after just a 
few years of
running at low luminosity if the signal is not observed.

At the LHC, the reach in the LFV coupling $\lambda_{\tau\mu}$ would be extended
beyond that expected at the TeVatron and also beyond that obtained from the muon 
$g-2$ data: a factor of 10 to 100 better depending on $\tan\beta$ and $\alpha$.
 
\vspace*{12pt}
\section*{Acknowledgments}
\label{sec:acknow}

This work is partially performed within the ATLAS collaboration and we thank 
collaboration members for helpful comments. We have used the physics analysis 
framework and tools which are the results of collaboration wide efforts. The 
authors would like to thank E.~Richter-W\c{a}s and G.~Azuelos for discussions 
and comments, and also to T.~Sj\"{o}strand for helpful correspondence.  

\section*{Appendix}

In this appendix we give the Lagrangian of the lepton flavor conserving and 
violating Yukawa couplings of the 2HDM type III using the  
results of \cite{Diaz:2001qb} with the addition of the charged Higgs part. The 
leptonic Yukawa Lagrangian reads:
\begin{equation}
-{\cal L}=\eta_{ij} {\bar l}^0_{iL} \varphi_1 l_{jR}^0 +\xi_{ij} {\bar 
l}_{iL}^0 \varphi_2 l_{jR}^0 + {\mathrm {h.c.}}
\label{eq:lagY}
\end{equation}
where $\varphi_{1,2}$ are the two Higgs doublets 
\begin{equation}
\varphi_1=\left(\matrix{\phi_1^+ \cr \phi_1^0 \cr} \right)~~~~~~~~~
\varphi_2=\left(\matrix{\phi_2^+ \cr \phi_2^0 \cr} \right)
\label{eq:higgses}
\end{equation}
with vacuum expectation values
\begin{equation}
\langle \varphi_1 \rangle_0 = \frac{1}{\sqrt{2}}\left(
\matrix{0 \cr v_1\cr} 
\right)~~~~~~~~~~
\langle \varphi_2 \rangle_0 = \frac{1}{\sqrt{2}}\left(
\matrix{0 \cr v_2 e^{i\theta}\cr} 
\right)\; ,
\end{equation}
where in the following we set $\theta=0$ and therefore we consider a CP 
conserving Higgs sector. The parameters $\eta_{ij}$ and $\xi_{ij}$ are non 
diagonal $3 \times 3$ matrices and $i$, $j$ are family indices. The neutral 
and charged mass eigenstates are related to the states of (\ref{eq:higgses}) 
by 
\begin{eqnarray}
\left(\matrix{\cos\beta 
& -\sin\beta\cr \sin \beta & \cos\beta\cr} \right) \left( \matrix{G_W^\pm\cr 
H^\pm\cr}\right) &=&\left(\matrix{\phi_1^\pm\cr \phi_2^\pm \cr} \right) 
\nonumber \\
\left(\matrix{\cos\beta & -\sin\beta\cr \sin \beta & \cos\beta\cr} \right)  
\left( \matrix{G_Z^0\cr A^0\cr}\right) &=&\sqrt{2} \left(\matrix{\Im 
\phi_1^0\cr \Im \phi_2^0 \cr} \right)\\
\left(\matrix{\cos\alpha & -\sin\alpha\cr \sin \alpha & \cos\alpha\cr} \right) 
\left( \matrix{H^0\cr h^0\cr}\right) &=&
\sqrt{2} \left(\matrix{\Re \phi_1^0 -v_1/\sqrt{2}\cr \Re \phi_2^0 
-v_2/\sqrt{2}\cr} \right) \nonumber
\end{eqnarray}
where $\Re \phi$ and $\Im \phi$ are the real and imaginary parts of the 
complex scalar fields $\phi$, $\tan \beta =v_2/v_1$, $\alpha$ is the CP-even 
neutral Higgs sector mixing angle, $G_Z^0$ and $G_W^\pm$ are the would-be 
Goldstone bosons of $Z$ and $W$ vector bosons, and $H^\pm$, $A^0$, $H^0$, 
$h^0$ are the physical Higgs bosons of the 2HDM. The Lagrangian 
(\ref{eq:lagY}) in terms of the mass eigenstates is obtained by a unitary 
transformation
\begin{equation}
l_{L,R}= V_{L,R} \, l^0_{L,R}
\end{equation}
and one can write the diagonal mass matrix for the three leptons
\begin{equation}
M_l^{\mathrm {diag}}= V_L \left( \frac{v_1}{\sqrt{2}} \eta + 
\frac{v_2}{\sqrt{2}} \xi \right) V^{\dag}_R
\end{equation} 
and either solve for $\xi$ (rotation of type a)
\begin{equation}
\xi=\frac{\sqrt{2}}{v_2} V_L^\dag M_l^{\mathrm {diag}} V_R -\frac{v_1}{v_2} 
\eta
\end{equation}
or for $\eta$ (rotation of type b)
\begin{equation}
\eta=\frac{\sqrt{2}}{v_1} V_L^\dag M_l^{\mathrm {diag}} V_R -\frac{v_2}{v_1} 
\xi \; .
\end{equation}
In terms of $\eta$ the leptonic Lagrangian reads (type a):
\begin{widetext}
\begin{eqnarray}
\label{eq:Yu1}
{\cal L}^a &=& -\frac{m_i}{v \sin \beta}{\bar l}_i  l_i 
(\cos \alpha \; h^0 + \sin \alpha \; H^0) 
-i \frac{m_i \cot \beta}{v} {\bar l}_i  \gamma_5 l_i A^0 
-\frac{m_i \cot \beta}{\sqrt{2} v}
{\bar n}_i  (1+\gamma_5) l_i H^+ \\
&+& \frac{1}{\sqrt{2} \sin\beta} {\bar l}_i \eta_{ij} l_j \left[ 
\cos(\alpha-\beta) h^0 + \sin(\alpha-\beta) H^0 \right]
+ \frac{i}{\sqrt{2} \sin\beta} {\bar l}_i \eta_{ij} \gamma_5 l_j A^0 +
\frac{1}{2\sin\beta}{\bar n}_i \eta_{ij} (1+\gamma_5) l_j H^+ 
+{\mathrm {h.c.}} \nonumber 
\end{eqnarray}
while in terms of $\xi$ the leptonic Lagrangian reads (type b):
\begin{eqnarray}
\label{eq:Yu2} 
{\cal L}^{b} &=& -\frac{m_i}{v \cos \beta}{\bar l}_i  l_i 
(\sin \alpha \; h^0 - \cos \alpha \; H^0) 
+i \frac{m_i \tan \beta}{v} {\bar l}_i  \gamma_5 l_i A^0 
+\frac{m_i \tan \beta}{\sqrt{2} v}
{\bar n}_i  (1+\gamma_5) l_i H^+  \\
&-& \frac{1}{\sqrt{2} \cos\beta} {\bar l}_i \xi_{ij} l_j \left[ 
\cos(\alpha-\beta) h^0+ \sin(\alpha-\beta) H^0  \right]
- \frac{i}{\sqrt{2} \cos\beta} {\bar l}_i \xi_{ij} \gamma_5 l_j A^0 -
\frac{1}{2 \cos\beta}{\bar n}_i \xi_{ij} (1+\gamma_5) l_j H^+ 
+{\mathrm {h.c.}}\nonumber
\end{eqnarray}
\end{widetext}
where $n$ is the neutrino field, $v=(\sqrt{2} \; G_F)^{-1/2}=246$~GeV is the 
SM vacuum expectation value, related to $v_1$ and $v_2$ by
\begin{equation}
v=\sqrt{v_1^2+v_2^2}\; .
\end{equation}
Note that the Lagrangian (\ref{eq:Yu1}) corresponds in the lepton flavor 
conserving part to 2HDM-I, while the Lagrangian (\ref{eq:Yu2}) to 2HDM-II.
\begin{table*}
\begin{center}
\begin{minipage}{0.6\linewidth} 
\caption{\label{tab:ffcoupl}The mixing angles for the neutral Higgs bosons 
decays to fermions; $u$ refers to up-type quarks, $d$ to down-type quarks and 
leptons.}
\end{minipage}
\vbox{\offinterlineskip 
\halign{&#& \strut\quad#\hfil\quad\cr  
\colrule
&  &&  MSSM   && 2HDM-III type a   && 2HDM-III type b   & \cr 
\colrule
& $\theta(h^0 \to u{\bar u})$ && ${\cos \alpha}/{\sin \beta}$  && ${\cos 
\alpha}/{\sin \beta}$ &&  ${\sin \alpha}/{\cos \beta}$ &\cr
& $\theta(h^0 \to d{\bar d})$ && ${\sin \alpha}/{\cos \beta}$  && ${\cos 
\alpha}/{\sin \beta}$ &&  ${\sin \alpha}/{\cos \beta}$ &\cr
& $\theta(h^0 \to \tau\mu)$ && 0 && ${\cos (\alpha-\beta)}/{(\sqrt{2} \sin 
\beta)}$ &&  ${\cos (\alpha-\beta)}/{(\sqrt{2}\cos \beta)}$ &\cr
& $\theta(H^0 \to u{\bar u})$ && ${\sin \alpha}/{\sin \beta}$  && ${\sin 
\alpha}/{\sin \beta}$ &&  ${\cos \alpha}/{\cos \beta}$ &\cr
& $\theta(H^0 \to d{\bar d})$ && ${\cos \alpha}/{\cos \beta}$  && ${\sin 
\alpha}/{\sin \beta}$ &&  ${\cos \alpha}/{\cos \beta}$ &\cr
& $\theta(H^0 \to \tau\mu)$ && 0 && ${\sin (\alpha-\beta)}/{(\sqrt{2} \sin 
\beta)}$ &&  ${\sin (\alpha-\beta)}/{(\sqrt{2}\cos \beta)}$ &\cr
& $\theta(A^0 \to u{\bar u})$ && ${\cot \beta}$  && ${\cot \beta}$ &&  
${\tan \beta}$ &\cr
& $\theta(A^0 \to d{\bar d})$ && ${\tan \beta}$  && ${\cot \beta}$ &&  
${\tan \beta}$ &\cr
& $\theta(A^0 \to \tau\mu)$ && 0 && $1/{(\sqrt{2} \sin \beta)}$ &&  
$1/{(\sqrt{2}\cos \beta)}$ &\cr
\colrule}}
\end{center}
\end{table*}
\begin{table*}
\begin{center}
\begin{minipage}{0.6\linewidth} 
\caption{\label{tab:cih}The coefficients $C^\phi_i$; $u$ refers to up-type 
quarks, $d$ to down-type quarks and leptons.}
\end{minipage}
\vbox{\offinterlineskip 
\halign{&#& \strut\quad#\hfil\quad\cr  
\colrule
&  && 2HDM-III type a   && 2HDM-III type b   & \cr 
\colrule
& $C^h_u$ && ${\cos \alpha}/{\sin \beta}$  && $-{\sin \alpha}/{\cos \beta}$ 
&\cr
& $C^h_d$ && ${\cos \alpha}/{\sin \beta}$  && $-{\sin \alpha}/{\cos \beta}$ 
&\cr
& $C^h_W$ && $\sin (\beta-\alpha)$  && $\sin (\beta-\alpha)$ &\cr
& $C^h_{H^\pm}$ && $g(h^0H^+H^-)$  && $g(h^0H^+H^-)$ &\cr
\colrule
& $C^H_u$ && ${\sin \alpha}/{\sin \beta}$  && ${\cos \alpha}/{\cos \beta}$ &\cr
& $C^H_d$ && ${\sin \alpha}/{\sin \beta}$  && ${\cos \alpha}/{\cos \beta}$ &\cr
& $C^H_W$ && $\cos (\beta-\alpha)$  && $\cos (\beta-\alpha)$ &\cr
& $C^H_{H^\pm}$ && $g(H^0H^+H^-)$  && $g(H^0H^+H^-)$ &\cr
\colrule
& $C^A_u$ && $-{\cot \beta}$  && ${\tan \beta}$ &\cr
& $C^A_d$ && ${\cot \beta}$  && $-{\tan \beta}$ &\cr
& $C^A_W$ && $0$  && $0$ &\cr
& $C^A_{H^\pm}$ && $0$  && $0$ &\cr
\colrule}}
\end{center}
\end{table*}

The couplings for lepton flavor conserving and violating Yukawa interactions 
$h_{ij}$ can be read directly from the Lagrangian (\ref{eq:Yu1}) and 
(\ref{eq:Yu2}). As an example the lepton flavor conserving charged Higgs 
coupling squared from (\ref{eq:Yu2}) (which is the same as in 2HDM-II) is
\begin{equation}
h^2_{\mu\nu}=\frac{m^2_\mu \tan^2 \beta}{2 v^2}=\frac{G_F \; m_\mu^2 \tan^2 
\beta}{\sqrt{2}}
\end{equation}
in agreement with the erratum in \cite{2HDM} (see the discussion there for a 
comment on other results in the literature).

We give in the following the complete expressions for the widths used in the 
analysis. For tree-level widths we do not give here loop contributions and 
threshold effects, but those effects are taken into account in the numerical 
calculation of the branching ratios.
The decays of a CP-even/odd neutral Higgs boson to a pair of fermions are:
\begin{widetext}
\begin{eqnarray}
\Gamma(H^0 \to l_i^+ l_j^-) &=& m_H 
\frac{N_c \lambda_{ij}^2}{8\pi}\frac{m_i m_j}{v^2} \, \theta^2(\alpha , \beta 
) \, \left( 1-\frac{(m_i +m_j)^2}{m_H^2} \right)^{3/2} \left( 1-\frac{(m_i 
-m_j)^2}{m_H^2} \right)^{1/2}, \\
\Gamma(A^0 \to l_i^+ l_j^-) &=& m_A 
\frac{N_c \lambda_{ij}^2}{8\pi}\frac{m_i m_j}{v^2} \, \theta^2(\alpha , \beta 
) \, \left( 1-\frac{(m_i +m_j)^2}{m_A^2} \right)^{1/2} \left( 1-\frac{(m_i 
-m_j)^2}{m_A^2} \right)^{3/2}, 
\end{eqnarray} 
\end{widetext}
where $N_c=3$ for quarks and $N_c=1$ for leptons. For a flavor conserving 
decay $\lambda_{ii}=1$ and $m_i=m_j$. $\theta (\alpha , \beta )$ is a function 
of the mixing parameters, given in Table~\ref{tab:ffcoupl}. The Higgs couplings 
to gauge bosons follow from gauge invariance and are therefore model independent. 
There are no tree-level couplings of vector boson pairs to the charged Higgs 
$H^\pm$ and to the CP-odd neutral Higgs boson $A$. 
For the neutral CP-even sector:
\begin{widetext}
\begin{eqnarray}
\Gamma(h^0 \to W^+W^-)&=&\frac{\sin^2(\beta-\alpha)}{16 \pi v^2 m_h} \left( 
m_h^4-4 m_h^2 m_W^2+ 12 m_W^4\right) \left(1-4 
\frac{m_W^2}{m_h^2}\right)^{1/2},\\ 
\Gamma(h^0 \to ZZ)&=&\frac{\sin^2(\beta-\alpha)}{32 \pi v^2 m_h} \left( 
m_h^4-4 m_h^2 m_Z^2+ 12 m_Z^4\right) \left(1-4 \frac{m_Z^2}{m_h^2}\right)^{1/2},
\end{eqnarray}
\end{widetext} 
and the corresponding expressions for $H^0$ can be obtained replacing 
$\sin^2(\beta-\alpha)$ with $\cos^2(\beta-\alpha)$. The loop-induced decays 
to $gg$ and $\gamma \gamma$ can be obtained from chapter 2 and appendix C of 
the first reference in \cite{2HDM} for the MSSM. For a generic neutral Higgs 
boson $\phi$ they are given by:
\begin{eqnarray}
\Gamma(\phi^0 \to gg)&=&\frac{\alpha_s^2}{128 \pi^3 v^2} m_{\phi}^3 \vert 
\sum_i J_i^\phi \vert^2,\\
\Gamma(\phi^0 \to \gamma \gamma)&=&\frac{\alpha_{em}^2}{256 \pi^3 v^2} 
m_{\phi}^3 \vert \sum_i I_i^\phi \vert^2,
\end{eqnarray}
where the sum over the index $i$ is limited to quarks for $gg$
\begin{equation}
J^h_q=C_q^\phi F_{1/2}(\tau_q)
\end{equation}
while it runs over fermions, $W$, $H^\pm$ for $\gamma\gamma$:
\begin{eqnarray}
I^\phi_f&=&N_c e^2 C^\phi_f F_{1/2}(\tau_f)\\
I^\phi_W&=&C^\phi_W F_1(\tau_W)\\
I^\phi_{H^\pm}&=&C^\phi_{H^\pm} F_0(\tau_{H^\pm}) \frac{m_W^2}{m_{H^\pm}^2}
\end{eqnarray}
where $\tau_i=4m_i^2/m^2_\phi$, $N_c=3$ for quarks, $N_c=1$ for leptons, $e$ 
is the electric charge in units of the charge of the electron, the functions 
$F$ are given by
\begin{eqnarray}
F_0&=&\tau\left[1-\tau g(\tau)\right]\\
F_{1/2}&=&-2\tau\left[ \delta+(1-\delta \tau)g(\tau)\right]\\
F_1&=&2+3\tau+3\tau(2-\tau)g(\tau)
\end{eqnarray}
where $\delta=1$ for $h^0$, $H^0$, and $\delta=0$ for $A$. The function 
$g(x)$ is
\begin{equation}
g(x)=\left\{ 
\begin{array}{lc}
{\mathrm{asin}}^2(\sqrt{1/x}) & x\ge 1 \\
-\frac{1}{4}\left[\log \frac{1+\sqrt{1-x}}{1-\sqrt{1-x}}-i\pi\right]&  x <1 
\end{array}
\right.
\end{equation}
The coefficients $C^\phi_i$ are given in Table~\ref{tab:cih} and the 
couplings $g(h^0H^+H^-)$, $g(H^0H^+H^-)$ of Table~\ref{tab:cih} can be
found in Appendix A of the last paper in~\cite{Haber}.


\begin{thebibliography}{99} 

\bibitem{2HDM}
J.F.~Gunion, H.E.~Haber, G.L.~Kane, S.~Dawson, The Higgs Hunter's Guide 
(Addison-Wesley, 
Reading, MA 1990), 
Erratum: arXiv:hep-ph/9302272;
%%CITATION = HEP-PH 9302272;%%
H.~E.~Haber and G.~L.~Kane,
%``The Search For Supersymmetry: Probing Physics Beyond The Standard Model,''
Phys.\ Rept.\  {\bf 117}, 75 (1985).
%%CITATION = PRPLC,117,75;%%

\bibitem{FCNC}
T.~P.~Cheng and M.~Sher,
Phys.\ Rev.\ D {\bf 35}, 3484 (1987);
%%CITATION = PHRVA,D35,3484;%%
A.~Antaramian, L.~J.~Hall and A.~Rasin,
%``Flavor changing interactions mediated by scalars at the weak scale,''
Phys.\ Rev.\ Lett.\  {\bf 69}, 1871 (1992)
[arXiv:hep-ph/9206205];
%%CITATION = HEP-PH 9206205;%%
L.~J.~Hall and S.~Weinberg,
%``Flavor changing scalar interactions,''
Phys.\ Rev.\ D {\bf 48}, 979 (1993)
[arXiv:hep-ph/9303241];
%%CITATION = HEP-PH 9303241;%%
M.~J.~Savage,
%``Constraining flavor changing neutral currents with B $\to$ mu+ mu-,''
Phys.\ Lett.\ B {\bf 266}, 135 (1991);
%%CITATION = PHLTA,B266,135;%%
M.~E.~Luke and M.~J.~Savage,
%``Flavor changing neutral currents in the Higgs sector and rare top decays,''
Phys.\ Lett.\ B {\bf 307}, 387 (1993)
[arXiv:hep-ph/9303249].
%%CITATION = HEP-PH 9303249;%%

\bibitem{Nilles:1984ge}
H.~P.~Nilles,
%``Supersymmetry, Supergravity And Particle Physics,''
Phys.\ Rept.\  {\bf 110}, 1 (1984).
%%CITATION = PRPLC,110,1;%%

\bibitem{giuratt}
G.~F.~Giudice and R.~Rattazzi,
%``Theories with gauge-mediated supersymmetry breaking,''
Phys.\ Rept.\  {\bf 322}, 419 (1999)
[arXiv:hep-ph/9801271].
%%CITATION = HEP-PH 9801271;%%

\bibitem{Barbieri:1996uv}
R.~Barbieri, G.~R.~Dvali and L.~J.~Hall,
%``Predictions From A U(2) Flavour Symmetry In Supersymmetric Theories,''
Phys.\ Lett.\ B {\bf 377}, 76 (1996)
[arXiv:hep-ph/9512388].
%%CITATION = HEP-PH 9512388;%%

\bibitem{Hall:1986dx}
L.~J.~Hall, V.~A.~Kostelecky and S.~Raby,
%``New Flavor Violations In Supergravity Models,''
Nucl.\ Phys.\ B {\bf 267}, 415 (1986).
%%CITATION = NUPHA,B267,415;%%

\bibitem{Borzumati:1986qx}
F.~Borzumati and A.~Masiero,
%``Large Muon And Electron Number Violations In Supergravity Theories,''
Phys.\ Rev.\ Lett.\  {\bf 57}, 961 (1986).
%%CITATION = PRLTA,57,961;%%

\bibitem{JELL} 
J.~Ellis, M.~E.~Gomez, G.~K.~Leontaris, S.~Sola and 
D.~V.~Nanopoulos, hep-ph/9911459.
%%CITATION = HEP-PH 9911459;%%

\bibitem{FENG} 
J.~L.Feng, Y.~Nir and Y.~Shadmi, Phys. Rev. D \textbf{61} 
(2000) 113005 [arXiv:hep-ph/9911370].
%%CITATION = HEP-PH 9911370;%%

\bibitem{HISA} 
J.~Hisano, T.~Moroi, K.~Tobe and M.~Yamagushi, Phys. Rev. D 
\textbf{53} (1996) 2442 [arXiv:hep-ph/9510309];
%%CITATION = HEP-PH 9510309;%%
Phys. Lett. B \textbf{391}, (1997) 341
[arXiv:hep-ph/9605296];
%%CITATION = HEP-PH 9605296;%%
J.~Hisano, D.Nomura and T.~Yanagida, Phys. Lett. B \textbf{437} (1998) 351
[arXiv:hep-ph/9711348];
%%CITATION = HEP-PH/9711348;%%
J.~Hisano and D.~Nomura, Phys. Rev. D \textbf{59} (1999) 116005
[arXiv:hep-ph/9810479];
%%CITATION = HEP-PH 9810479;%%
J.~Hisano, M.~M.~Nojiri, Y.~Shimizu and M.~Tanaka, Phys. Rev. D \textbf{60} 
(1999) 055008 [arXiv:hep-ph/9808410];
%%CITATION = HEP-PH 9808410;%%
Phys. Rev. D \textbf{58}, (1998) 116010 [arXiv:hep-ph/9805367];
%%CITATION = HEP-PH 9805367;%%
J.~Hisano, hep-ph/9806222;
%%CITATION = HEP-PH/9806222;%% 
J.~Hisano, hep-ph/9906312;
%%CITATION = HEP-PH/9906312;%% 
J.~Hisano and K.~Kurosawa, hep-ph/0004061.
%%CITATION = HEP-PH 0004061;%%

\bibitem{Huitu}
M.~Chaichian and K.~Huitu,
%``Constraints on $R$-parity violating interactions from $\mu\to e\gamma$,''
Phys.\ Lett.\ B {\bf 384}, 157 (1996)
[arXiv:hep-ph/9603412];
%%CITATION = HEP-PH 9603412;%%
K.~Huitu, J.~Maalampi, M.~Raidal and A.~Santamaria,
%``New constraints on R-parity violation from mu e conversion in nuclei,''
Phys.\ Lett.\ B {\bf 430}, 355 (1998)
[arXiv:hep-ph/9712249].
%%CITATION = HEP-PH 9712249;%%

\bibitem{GLAS-WEIN}
S.~L.~Glashow and S.~Weinberg,
%``Natural Conservation Laws For Neutral Currents,''
Phys.\ Rev.\ D {\bf 15}, 1958 (1977).
%%CITATION = PHRVA,D15,1958;%%

\bibitem{BWOS}
D.~Bowser-Chao, K.~Cheung and W.-Y.~Keung, Phys. Rev. D \textbf{59}, 115006 
(1999) [arXiv:hep-ph/9811235].
%%CITATION = HEP-PH 9811235;%%

\bibitem{BCH}
G.~Buchalla, A.~Buras and M.~Lautenbacher, Rev. Mod. Phys. \textbf{68}, 1125 
(1996) [arXiv:hep-ph/9512380];
%%CITATION = HEP-PH 9512380;%% 
A.~J.~Buras, M.~Misiak, M.~Munz and S.~Pokorski, Nucl. Phys. \textbf{B424}, 
374 (1994) [arXiv:hep-ph/9311345]. 
%%CITATION = HEP-PH 9311345;%%

\bibitem{CHET}
K.~Chetyrkin, M.~Misiak and M.~Munz, Phys. Lett. \textbf{B400}, 206 (1997); 
Erratum-ibid. \textbf{B425}, 414 (1998) [arXiv:hep-ph/9612313];
%%CITATION = HEP-PH 9612313;%% 
M.Ciuchini, G.Degrassi, P.~Gambino and G.~F.~Giudice, Nucl. 
Phys. \textbf{B527}, 21 (1998) [arXiv:hep-ph/9710335]; 
%%CITATION = HEP-PH 9710335;%%
A.~Kagan and M.Neubert, Eur. Phys. J. C \textbf{7}, 5 (1999).
%%CITATION = NONE;%%

\bibitem{CLEO}
M.~S.~Alam, CLEO Collaboration, Phys. Rev. Lett. \textbf{74}, 2885 (1995); 
R.~Briere, in Proceedings of ICHEP98, Vancouver, Canada 1998, CLEO-CONF-98-17; 
and in talk by J. Alexander, \textit{ibid}.
%%CITATION = PRLTA,74,2885;%%

\bibitem{ALEPH}
R.~Barate, ALEPH Collaboration, Phys. Lett. B \textbf{429}, 169 (1998).
%%CITATION = NONE;%%

\bibitem{HP-BOUND}
J.~L.~Hewett,
%``Can b $\to$ s gamma close the supersymmetric Higgs production window?,''
Phys.\ Rev.\ Lett.\  {\bf 70}, 1045 (1993)
[arXiv:hep-ph/9211256];
%%CITATION = HEP-PH 9211256;%%
V.~D.~Barger, M.~S.~Berger and R.~J.~Phillips,
Phys.\ Rev.\ Lett.\  {\bf 70}, 1368 (1993) [arXiv:hep-ph/9211260];
%%CITATION = HEP-PH 9211260;%%
F.~M.~Borzumati and C.~Greub, Phys. Rev. D \textbf{58}, 074004 (1998) 
[arXiv:hep-ph/9802391];
%%CITATION = HEP-PH 9802391;%% 
J.~A.~Coarasa, J.~Guasch, J.~Sola and W.~Hollik,
Phys.\ Lett.\ B {\bf 442}, 326 (1998)
[arXiv:hep-ph/9808278];
%%CITATION = HEP-PH 9808278;%%
F.~M.~Borzumati and C.~Greub, Phys.\ Rev.\ D {\bf 59}, 057501 (1999)
[arXiv:hep-ph/9809438];
%%CITATION = HEP-PH 9809438;%%
G.~Degrassi, P.~Gambino and G.~F.~Giudice, JHEP {\bf 0012}, 009 (2000)
[arXiv:hep-ph/0009337];
%%CITATION = HEP-PH 0009337;%%
P.~Gambino and M.~Misiak,
%``Quark mass effects in anti-B $\to$ X/s gamma,''
Nucl.\ Phys.\ B {\bf 611}, 338 (2001)
[arXiv:hep-ph/0104034].
%%CITATION = HEP-PH 0104034;%%

\bibitem{ATWO}
A.~Atwood, L.~Reina and A.~Soni, Phys. Rev. D \textbf{55}, 3156 (1997), 
[arXiv:hep-ph/9609279].
%%CITATION = HEP-PH 9609279;%%

\bibitem{DIAZ}
J.~L.~Diaz Cruz, J.~J.~Godina Nava and G.~Lopez Castro,
%``Low-energy effects of Charged Higgs with general Yukawa couplings,''
Phys.\ Rev.\ D {\bf 51}, 5263 (1995)
[arXiv:hep-ph/9509229].
%%CITATION = HEP-PH 9509229;%%

\bibitem{SHER3} 
M.~Sher and Y.~Yuan,
%``Rare B decays, rare tau decays and grand unification,''
Phys.\ Rev.\ D {\bf 44}, 1461 (1991).
%%CITATION = PHRVA,D44,1461;%%

\bibitem{BOUND}
S.~Nie and M.~Sher,
Phys.\ Rev.\ D {\bf 58}, 097701 (1998)
[arXiv:hep-ph/9805376].
%%CITATION = HEP-PH 9805376;%%

\bibitem{Brown:2001mg}
H.~N.~Brown {\it et al.}  [Muon g-2 Collaboration],
%``Precise measurement of the positive muon anomalous magnetic moment,''
Phys.\ Rev.\ Lett.\  {\bf 86}, 2227 (2001)
[arXiv:hep-ex/0102017];
%%CITATION = HEP-EX 0102017;%%
G.~W.~Bennett {\it et al.}  [Muon g-2 Collaboration],
%``Measurement of the Positive Muon Anomalous Magnetic Moment to 0.7 ppm,''
arXiv:hep-ex/0208001.
%%CITATION = HEP-EX 0208001;%%

\bibitem{llh}
M.~Knecht and A.~Nyffeler, arXiv:hep-ph/0111058;
%%CITATION = HEP-PH 0111058;%%
M.~Knecht, A.~Nyffeler, M.~Perrottet and E.~De Rafael,
Phys.\ Rev.\ Lett.\  {\bf 88}, 071802 (2002) [arXiv:hep-ph/0111059];
%%CITATION = HEP-PH 0111059;%%
I.~Blokland, A.~Czarnecki and K.~Melnikov,
Phys.\ Rev.\ Lett.\  {\bf 88}, 071803 (2002) [arXiv:hep-ph/0112117];
%%CITATION = HEP-PH 0112117;%%
J.~Bijnens, E.~Pallante and J.~Prades, Nucl.\ Phys.\ B {\bf 626}, 410 (2002)
[arXiv:hep-ph/0112255].
%%CITATION = HEP-PH 0112255;%%

\bibitem{Ramsey-Musolf:2002cy}
M.~Ramsey-Musolf and M.~B.~Wise,
arXiv:hep-ph/0201297.
%%CITATION = HEP-PH 0201297;%%

\bibitem{Leveille:1978rc}
J.~P.~Leveille, Nucl.\ Phys.\ B {\bf 137}, 63 (1978).
%%CITATION = NUPHA,B137,63;%%

\bibitem{Haber}
H.~E.~Haber and R.~Hempfling,
Phys.\ Rev.\ Lett.\  {\bf 66}, 1815 (1991);
%%CITATION = PRLTA,66,1815;%%
J.~R.~Ellis, G.~Ridolfi and F.~Zwirner,
%``Radiative corrections to the masses of supersymmetric Higgs bosons,''
Phys.\ Lett.\ B {\bf 257}, 83 (1991);
Y.~Okada, M.~Yamaguchi and T.~Yanagida,
Prog.\ Theor.\ Phys.\  {\bf 85}, 1 (1991);
%%CITATION = PTPKA,85,1;%%
%%CITATION = PHLTA,B257,83;%%
H.~E.~Haber,
%``Higgs boson masses and couplings in the minimal supersymmetric model,''
arXiv:hep-ph/9707213.
%%CITATION = HEP-PH 9707213;%%

\bibitem{AGUI} 
J.~A.~Aguilar-Saavedra and G.~C.~Branco,
%``Probing top flavour-changing neutral scalar couplings at the CERN LHC,''
Phys.\ Lett.\ B {\bf 495}, 347 (2000)
[arXiv:hep-ph/0004190].
%%CITATION = HEP-PH 0004190;%%

\bibitem{SANT} 
S.~Bejar, J.~Guasch and J.~Sola,
%``FCNC top quark decays beyond the standard model,''
in {\it Proc. of the 5th International Symposium on Radiative Corrections 
(RADCOR 2000) } ed. Howard E. Haber,
arXiv:hep-ph/0101294.
%%CITATION = HEP-PH 0101294;%%

\bibitem{Han-Sher} 
T.~Han, J.~Jiang and M.~Sher,
%``Search for t $\to$ c h at e+ e- linear colliders,''
Phys.\ Lett.\ B {\bf 516}, 337 (2001)
[arXiv:hep-ph/0106277].
%%CITATION = HEP-PH 0106277;%%

\bibitem{KAMIO}
Y.~Fukuda {\it et al.}  [SuperKamiokande Collaboration], 
Phys.\ Rev.\ Lett.\  {\bf 82}, 2644 (1999) 
[arXiv:hep-ex/9812014]; 
%%CITATION = HEP-EX 9812014;%% 
{\bf 85}, 3999 (2000) 
[arXiv:hep-ex/0009001]; 
%%CITATION = HEP-EX 0009001;%% 
{\bf 86}, 5656 (2001) 
[arXiv:hep-ex/0103033].
%%CITATION = HEP-EX 0103033;%% 

\bibitem{SERI} 
L.~Serin and R.~Stroynowski, ATLAS Internal Note, 
ATL-PHYS-97-114 (1997).
%%CITATION =NONE;%% 

\bibitem{HINC} 
I.~Hinchliffe and F.~E.~Paige,
%``Lepton flavor violation at the LHC,''
Phys.\ Rev.\ D {\bf 63}, 115006 (2001)
[arXiv:hep-ph/0010086].
%%CITATION = HEP-PH 0010086;%%

\bibitem{TOSC} 
J.~L.~Diaz-Cruz and J.~J.~Toscano,
Phys.\ Rev.\ D {\bf 62}, 116005 (2000)
[arXiv:hep-ph/9910233].
%%CITATION = HEP-PH 9910233;%%

\bibitem{SHER2} 
M.~Sher,
%``Scalar mediated FCNC at the first muon collider,''
Phys.\ Lett.\ B {\bf 487}, 151 (2000) [arXiv:hep-ph/0006159].
%%CITATION = HEP-PH 0006159;%%

\bibitem{Cotti:2001fm}
U.~Cotti, L.~Diaz-Cruz, C.~Pagliarone and E.~Vataga,
in {\it Proc. of the APS/DPF/DPB Summer Study on the Future of Particle 
Physics (Snowmass 2001) } ed. R.~Davidson and C.~Quigg,
arXiv:hep-ph/0111236.
%%CITATION = HEP-PH 0111236;%%

\bibitem{HDECAY} 
A.~Djouadi, J.~Kalinowski and M.~Spira,
Comput.\ Phys.\ Commun.\  {\bf 108}, 56 (1998)
[arXiv:hep-ph/9704448].
%%CITATION = HEP-PH 9704448;%%

\bibitem{PYTHIA}  
T.~Sj\"{o}strand,
%``High-energy physics event generation with PYTHIA 5.7 and JETSET 7.4,''
Comput.\ Phys.\ Commun.\  {\bf 82}, 74 (1994);
%%CITATION = CPHCB,82,74;%%
T.~Sj\"{o}strand, P.~Eden, C.~Friberg, L.~Lonnblad, G.~Miu, S.~Mrenna and 
E.~Norrbin,
%``High-energy-physics event generation with PYTHIA 6.1,''
Comput.\ Phys.\ Commun.\  {\bf 135}, 238 (2001)
[arXiv:hep-ph/0010017];
%%CITATION = HEP-PH 0010017;%%
T.~Sj\"{o}strand, L.~Lonnblad and S.~Mrenna,
%``PYTHIA 6.2: Physics and manual,''
arXiv:hep-ph/0108264.
%%CITATION = HEP-PH 0108264;%%
 
\bibitem{CTEQ}  
H.~L.~Lai {\it et al.},
Phys.\ Rev.\ D {\bf 55}, 1280 (1997)
[arXiv:hep-ph/9606399].
%%CITATION = HEP-PH 9606399;%%

\bibitem{ATLFAST}  
E.~Richter-W\c{a}s, D.~Froidevaux and L.~Poggioli, ATLAS Internal Note, 
ATL-PHYS-98-131, (1998). 
%%CITATION = NONE;%% 

\bibitem{HIGL} M.~Spira, hep-ph/9510347.
%%CITATION = HEP-PH 9510347;%% 

\bibitem{BALA} 
S.~Balatenychev et al., 'Theoretical Developments', 
in~\cite{LesHouches2001}, p.4; and the references therein.

\bibitem{CATA} 
S.~Catani, D.~de Florian and M.~Grazzini,
JHEP {\bf 0201}, 015 (2002)
[arXiv:hep-ph/0111164].
%%CITATION = HEP-PH 0111164;%%

\bibitem{FRIX} 
S.~Frixione, P.~Nason and G.~Ridolfi,
%``Strong corrections to W Z production at hadron colliders,''
Nucl.\ Phys.\ B {\bf 383}, 3 (1992);
%%CITATION = NUPHA,B383,3;%%
S.~Frixione, Nucl.\ Phys.\ B {\bf 410}, 280 (1993);
%%CITATION = NUPHA,B410,280;%%
U.~Baur, T.~Han and J.~Ohnemus, Phys.\ Rev.\ D {\bf 48}, 5140 (1993)
[arXiv:hep-ph/9305314];
%%CITATION = HEP-PH 9305314;%%
Phys.\ Rev.\ D {\bf 51}, 3381 (1995) [arXiv:hep-ph/9410266];
%%CITATION = HEP-PH 9410266;%%
Phys.\ Rev.\ D {\bf 53}, 1098 (1996) [arXiv:hep-ph/9507336];
%%CITATION = HEP-PH 9507336;%%
Phys.\ Rev.\ D {\bf 57}, 2823 (1998) [arXiv:hep-ph/9710416].
%%CITATION = HEP-PH 9710416;%%

\bibitem{DIXO}
L.~J.~Dixon, Z.~Kunszt and A.~Signer,
Phys.\ Rev.\ D {\bf 60}, 114037 (1999)
[arXiv:hep-ph/9907305].
%%CITATION = HEP-PH 9907305;%%

\bibitem{CAMP} 
J.~Campbell and R.~K.~Ellis, 
{\url {http://theory.fnal.gov/people/ellis/Programs/mcfm.html}},
arXiv:hep-ph/0202176.
%%CITATION = HEP-PH 0202176;%%

\bibitem{TAUOLA} 
S.~Jadach, Z.~W\c{a}s, R.~Decker and J.~H.~K\"{u}hn, 
%``The tau decay library TAUOLA: Version 2.4,'' 
Comput.\ Phys.\ Commun.\  {\bf 76}, 361 (1993); 
%%CITATION = CPHCB,76,361;%% 
M.~Jezabek, Z.~W\c{a}s, S.~Jadach and J.~H.~K\"{u}hn, 
Comput.\ Phys.\ Commun.\  {\bf 70}, 69 (1992); 
%%CITATION = CPHCB,70,69;%% 
S.~Jadach, J.~H.~K\"{u}hn and Z.~W\c{a}s, 
Comput.\ Phys.\ Commun.\  {\bf 64}, 275 (1990). 
%%CITATION = CPHCB,64,275;%% 

\bibitem{HAN} 
T.~Han and D.~Marfatia,
%``h $\to$ mu tau at hadron colliders,''
Phys.\ Rev.\ Lett.\  {\bf 86}, 1442 (2001)
[arXiv:hep-ph/0008141].
%%CITATION = HEP-PH 0008141;%%

\bibitem{Rainwater} 
D.~Rainwater, D.~Zeppenfeld and K.~Hagiwara,
%``Searching for H $\to$ tau tau in weak boson fusion at the LHC,''
Phys.\ Rev.\ D {\bf 59}, 014037 (1999)
[arXiv:hep-ph/9808468];
%%CITATION = HEP-PH 9808468;%%
L. Di Lella, Proceedings of the large Hadron Collider Workshop, Aachen, Vol 
II, p.183, 1990, edited by G. Jarlskog and D.~Rein, CERN 90-10/ECFA 90-133.
%%CITATION = NONE;%%

\bibitem{CAVA} 
ATLAS Collaboration, 'ATLAS Detector and Physics Performance 
Technical Design Report', CERN/LHCC/99-15, 742 (1999); 
%%CITATION = NONE;%%
D.~Cavalli et al., ATLAS Internal Note ATL-PHYS-94-051; 
%%CITATION = NONE;%%
D.~Cavalli et al., 'Study of the MSSM channel $A/H \to \tau\tau$ at the LHC', 
in~\cite{LesHouches2001}, p.67.

\bibitem{RITVA} 
R.~Kinnunen and D.~Denegri, CMS NOTE 1999/037, 
arXiv:hep-ph/9907291;
%%CITATION = HEP-PH 9907291;%%
A.~Nikitenko, S.~Kunori and R.~Kinnunen, CMS Note 2001/040; 
%%CITATION = NONE;%%
D.~Denegri et al., CMS NOTE 2001/032, arXiv:hep-ph/0112045.
%%CITATION = HEP-PH 0112045;%%

\bibitem{Diaz:2001qb}
R.~A.~Diaz, R.~Martinez and J.~Alexis Rodriguez,
%``Lepton flavor violation in the two Higgs doublet model type III,''
Phys.\ Rev.\ D {\bf 63}, 095007 (2001);
%%CITATION = PHRVA,D63,095007;%%
Phys.\ Rev.\ D {\bf 64}, 033004 (2001) [arXiv:hep-ph/0103050].
%%CITATION = HEP-PH 0103050;%%

\bibitem{LesHouches2001} 
D.~Cavalli {\it et al.}, ``The Higgs working group: Summary report,''
arXiv:hep-ph/0203056.
%%CITATION = HEP-PH 0203056;%%


\end{thebibliography}
\end{document}